\documentclass[12pt]{article}
\pdfoutput=1
\usepackage{hyperref}
\usepackage{textcomp}
\usepackage[DIV13]{typearea}
\usepackage{amsmath, amsthm, amssymb, mathtools,empheq,latexsym,dsfont}
\usepackage{slashed, simplewick}
\usepackage[utf8]{inputenc}
\usepackage{graphicx,float,placeins}
\usepackage{makeidx}
\usepackage[font=small,labelfont=bf]{caption}
\usepackage{nicefrac}
\usepackage{subfigure}
\usepackage{array, bigdelim,multirow,multicol}
\usepackage[usenames]{color}
\usepackage[dvipsnames]{xcolor}
\usepackage{comment}

\usepackage{cite}

\graphicspath{{immagini/}}
 \def\cB{{\cal B}}
\def\cC{{\cal C}}

\def\cL{{\cal L}}

\newcommand{ \mysmall}[1]{\scriptscriptstyle #1}
\newcommand{\ba} {\begin{eqnarray}}
\newcommand{\ea} {\end{eqnarray}}
\newcommand{\nn}{\nonumber}
\newcommand{\be}{\begin{equation}}
\newcommand{\ee}{\end{equation}}
\newcommand{\bea}{\begin{eqnarray}}
\newcommand{\eea}{\end{eqnarray}}

\newcommand{\arXhref}[1]{\href{http://arxiv.org/abs/#1}{#1}}
\newcommand{\mL}{ \mysmall L}
\newcommand{\mR}{ \mysmall R}

\newcommand{\eftQF}[3]{\big[Q_{#1}^{#2}\big]_{#3}}

\newcommand{\RK} {R^{\mu/e}_K}
\newcommand{\RKs} {R^{\mu/e}_{K^\ast}}
\newcommand{\RKss} {R^{\mu/e}_{K^{(\ast)}}}

\begin{document}
%%%%%%%%%%%%%%%%%%%%%%%%%%%%%%%%%%%%%%%%%%%%%%%%%%%%%%
 \unitlength = 1mm

%\begin{titlepage}
%\vspace*{-1cm}
%\phantom{hep-ph/***} 
%\flushright
\qquad\qquad\qquad\qquad\qquad\qquad\qquad\qquad\qquad\qquad\qquad\qquad\qquad\qquad\hfil{ZU-TH-09/18}\\

\setlength{\extrarowheight}{0.2 cm}

\thispagestyle{empty}

\bigskip

\vskip 1cm

\begin{center}
\vspace{1.5cm}
    {\Large\bf Low-energy Effects of Lepton}\\[0.5 cm] {\Large\bf Flavour Universality Violation}\\[1cm]
%    {\Large\bf\red{[Draft 11.12.2017]}}\\[1cm]
    {\bf Claudia Cornella$^{a}$, Ferruccio Feruglio$^{b}$, Paride Paradisi$^{b}$}     \\[0.5cm]
  {\em $(a)$  Physik-Institut, Universit\"at Z\"urich, CH-8057 Z\"urich, Switzerland}\\   
     {\em $(b)$  Dipartimento di Fisica e Astronomia `G.~Galilei', Universit\`a di Padova\\
INFN, Sezione di Padova, Via Marzolo~8, I-35131 Padua, Italy}  \\[1.0cm]
\end{center}

\centerline{\large\bf Abstract}
\begin{quote}
\indent
The persisting anomalous data in semileptonic B-decays point towards New Physics models exhibiting large sources of 
Lepton Flavour Universality Violation. In this work we generalise previous studies by considering frameworks which include an 
enlarged set of semileptonic four-fermion operators invariant under the SM gauge group, with New Physics affecting mainly the 
third generation. We derive the low-energy effective Lagrangian including the leading electroweak corrections, mandatory to 
obtain reliable predictions. As a particularly interesting case, we analyse the scenario where the dominant New Physics effects 
are encoded in the Wilson coefficient $C_9$, as favoured by global fit analyses of $b\to s$ data. We find that also in this case 
the stringent experimental bounds on $Z$-pole observables and $\tau$ decays challenge a simultaneous explanation of charged 
and neutral-current non-standard data. 
\end{quote}

%\end{titlepage}

\newpage 

%\tableofcontents
%%%%%%%%%%%%%%%%%%%%%%%%%%%%%%%%%%%%%%%%%%%%%%%%%%%%%%%%%%%%%
\section{Introduction}
%%%%%%%%%%%%%%%%%%%%%%%%%%%%%%%%%%%%%%%%%%%%%%%%%%%%%%%%%%%%%

In the last few years, various experimental collaborations observed indications of Lepton Flavour Universality Violation (LFUV) 
in semileptonic $B$ decays. Although such indications are not yet conclusive,  the overall pattern of deviations from the Standard Model (SM) 
predictions is very coherent. The anomalous data refer to i) charged-current transitions $b\to c \ell \bar\nu$ with $\tau/e$ and $\tau/\mu$ LFUV~\cite{Lees:2013uzd,Aaij:2015yra,Hirose:2016wfn,Aaij:2017deq} and ii) neutral-current transitions $b\to s \ell \bar{\ell}$ with $\mu/e$ LFUV~\cite{Aaij:2014ora,Aaij:2017vbb}. Interestingly enough, global fit analyses for the angular distributions of the $B^0 \to K^{*0} \mu^+ \mu^-$ 
decay reported anomalies which are consistent with LFUV data~\cite{Altmannshofer:2014rta,Descotes-Genon:2015uva,Hiller:2014yaa}.\\
From a theoretical point of view, it would be desirable to explain both the charged- and neutral-current anomalies within a coherent extension 
of the SM~\cite{Greljo:2015mma,Falkowski:2015zwa,Barbieri:2015yvd,Barbieri:2016las,DiLuzio:2017vat,Calibbi:2017qbu,Bordone:2017bld,Barbieri:2017tuq,Becirevic:2016yqi}.
A first step towards this goal is represented by an effective theory where the effects of New Physics (NP) are described by
four-fermion operators involving left-handed currents, $(\bar{s}_L\gamma_\mu b_L)(\bar{\mu}_L\gamma_\mu \mu_L)$ and
$(\bar{c}_L\gamma_\mu b_L)(\bar{\tau}_L\gamma_\mu \nu_L)$, which are related by the $SU(2)_L$ gauge symmetry~\cite{Bhattacharya:2014wla,Calibbi:2015kma}. 
A crucial ingredient of such a theory requires that NP couples much more strongly to the third generation than to the first two, 
since $(\bar{c}_L\gamma_\mu b_L)(\bar{\tau}_L\gamma_\mu \nu_L)$ is induced already at the tree level in the SM while
$(\bar{s}_L\gamma_\mu b_L)(\bar{\mu}_L\gamma_\mu \mu_L)$ arises only at loop-level.
The latter requirement is realized, for instance, if NP is coupled only to the third fermion generation in the interaction basis. 
Couplings to lighter generations are generated after electroweak symmetry breaking by the misalignment between the mass 
and the interaction bases through small flavour mixing angles~\cite{Glashow:2014iga}. 

Hence, a minimal framework addressing the B-anomalies consists of an effective Lagrangian defined above the electroweak scale and 
containing gauge-invariant semileptonic operators involving purely left-handed fermions of the third generation. Assuming such starting point, in~\cite{Feruglio:2016gvd,Feruglio:2017rjo} the low-energy effective Lagrangian including leading electroweak corrections was derived.
The most striking effects found were large corrections to the leptonic couplings of the $W$ and $Z$ vector bosons and the generation of 
a purely leptonic effective Lagrangian. The resulting LFUV in $Z$ and $\tau$ decays and $\tau$ Lepton Flavour Violating (LFV) 
contributions turned out to challenge a simultaneous explanation of charged- and neutral-current anomalies. 
Although this conclusion applies under certain assumptions, our main message was that including electroweak corrections is mandatory 
when addressing the B-anomalies with NP at the TeV scale. 
Another important challenge that one has to face is the lack of signals in direct production at LHC of any mediators responsible 
of the four-fermion interactions invoked to explain the B-anomalies~\cite{Faroughy:2016osc,Greljo:2017vvb}.

In this paper we make a step forward compared to~\cite{Feruglio:2016gvd,Feruglio:2017rjo}. In particular, we consider both
purely left-handed operators $(V-A)\times(V-A)$ as well as operators with right-handed currents of 
the form $(V + A)\times(V + A)$ and $(V \pm A)\times(V \mp A)$.
This effort is justified by the fact that many NP models, proposed to accommodate B-anomalies, exhibit the operators 
considered here 
%\red{(add references?)} 
\footnote{We do not consider operators of scalar or tensor type.
The former are severely constrained by the $B_c$ lifetime through the enhancement of the $B_c^-\to \tau^- \bar{\nu}$ channel \cite{Alonso:2016oyd}.
Renormalization of scalar and tensor operators, their strong mixing and the impact on phenomenology has been recently analysed in
 ref.~\cite{Gonzalez-Alonso:2017iyc}.}.
Moreover, as we will discuss in the following, such enlarged operator basis will allow us to consider 
one of the most favoured solutions to the neutral-current anomalies, with dominant NP effects encoded in the 
low-energy Wilson coefficient $C_9$~\cite{Altmannshofer:2014rta,Descotes-Genon:2015uva,Hiller:2014yaa}.

The paper is organised as follows. In section 2, we present the theoretical framework and construct the low-energy 
effective Lagrangian including electroweak corrections in the leading logarithm approximation. 
In section 3, we examine the phenomenological implications of our setup, discussing both tree-level and loop-induced 
low-energy observables.
In section 4, we focus on the scenario where the dominant NP effects are encoded in the Wilson coefficient $C_9$,
providing a numerical analysis. Our conclusions are presented in section 5.

%%%%%%%%%%%%%%%%%%%%%%%%%%%%%%%%%%%%%%%
\section{Theoretical framework}
\label{flavour}
%%%%%%%%%%%%%%%%%%%%%%%%%%%%%%%%%%%%%%%

We assume that strong and electroweak interactions at the scale $\Lambda \gg m_W$ are described by the effective Lagrangian
\be
\cL=\cL_{\rm SM}+\cL_{\rm NP}^{\mysmall 0}\, ,
\ee
where the NP contribution is given by
\be
\label{eq:NP_Lambda}
\cL_{\rm NP}^{\mysmall 0} = \frac{1}{\Lambda^2} \left( C_1 [Q_{\ell q}^{\mysmall(1)} ]_{\mysmall 3333} + C_3 [Q_{\ell q}^{\mysmall(3)} ]_{\mysmall 3333} + C_4 [ Q_{\ell d} ]_{\mysmall 3333}+ C_5 [ Q_{ed} ]_{\mysmall 3333} + C_6 [ Q_{qe} ]_{\mysmall 3333} \right) \, 
\ee
and the semileptonic operators $Q_i$ are defined in table \ref{tab:rge_flow}, where primed fields indicate fields in the interaction basis. 
We denote the Wilson coefficients at the scale $\Lambda$ by 
$C_1=[\cC_{\ell q}^{\mysmall(1)}(\Lambda)]_{\mysmall 3333}$, $C_3=[\cC_{\ell q}^{\mysmall(3)}(\Lambda)]_{\mysmall 3333}$ and so on.
Notice that (\ref{eq:NP_Lambda}) assumes that NP couples only to third generation fermions. Couplings to light generations will arise when
switching from the interaction to the mass basis after electroweak symmetry breaking, as we will describe shortly. Such an assumption is motivated by the need of generating a hierarchy between NP effects in charged- and neutral-current semileptonic B-decays, as suggested by experimental data.
We move to the mass basis, denoted by unprimed fields, by means of the unitary transformations
\begin{align}
\begin{aligned}
\label{eq:int_to_mass}
u'_{\mL} &= V_u u_{\mL} &\qquad d'_{\mL} =& V_d d_{\mL} &\qquad \ell^\prime_{\mL} &= V_e \ell_{\mL} , \\
u'_{\mR} &= R_u u_{\mR} &\qquad d'_{\mR} =& R_d d_{\mR} &\qquad e'_{\mR} &= R_e e_{\mR} \, ,
\end{aligned}
\end{align}
where we work in the approximation of massless neutrinos. To keep track of the flavour structure of the Lagrangian, we define the following matrices in flavour space
\begin{align}
\begin{aligned}
\lambda^u_{ij}&=V_{u3i}^*V_{u3j} &\qquad \lambda^d_{ij}&=V_{d3i}^*V_{d3j} &\qquad \lambda^e_{ij} &=V_{e3i}^*V_{e3j} &\qquad \lambda^{ud}_{ij}&=V_{u3i}^*V_{d3j} \,  \\
\Gamma^d_{ij}&=R_{d3i}^*R_{d3j} &\qquad \Gamma^e_{ij}&=R_{e3i}^*R_{e3j}  \,,
\label{eq:lambda}
\end{aligned}
\end{align}
where $\lambda$ and $\Gamma$ are both projectors with trace equal to one, and the $\lambda$ matrices are related by $\lambda^u=V_{\rm \mysmall CKM} \lambda^d V_{\rm \mysmall CKM}^{\dagger}$ and $\lambda^{ud}=V_{\rm \mysmall CKM} \lambda^d$, $V_{\rm \mysmall CKM}=V_u^\dagger V_d$ being the quark mixing matrix. Hereafter we will omit the subscript CKM for simplicity.
In the mass basis the Lagrangian ${\cal L}_{\rm \mysmall NP}^{\mysmall 0}$ reads:
\begin{align}
\begin{aligned}
{\cal L}_{\rm \mysmall NP}^{\mysmall 0} &= \frac{1}{\Lambda^2} \Big[ (C_1-C_3) (\bar e_{\mL} \gamma^{\mu} \lambda_e e_{\mL})(\bar u_{\mL} \gamma_{\mu} \lambda_u u_{\mL}) + (C_1+C_3) (\bar e_{\mL} \gamma^{\mu} \lambda_e e_{\mL})(\bar d_{\mL} \gamma_{\mu} \lambda_d d_{\mL})  \\[0.1 cm]
&+ (C_1+C_3) (\bar \nu_{\mL} \gamma^{\mu} \lambda_e \nu_{\mL})(\bar u_{\mL} \gamma_{\mu} \lambda_d u_{\mL})+(C_1-C_3) (\bar \nu_{\mL} \gamma^{\mu} \lambda_e \nu_{\mL})(\bar d_{\mL} \gamma_{\mu} \lambda_d d_{\mL})  \\[0.2 cm]
&+\left( 2C_3 (\bar e_{\mL} \gamma^{\mu} \lambda_e \nu_{\mL})(\bar u_{\mL} \gamma_{\mu} \lambda_{ud} d_{\mL}) + {\rm h.c.}\right)+ C_5  (\bar e_{\mR} \gamma^{\mu} \Gamma_e e_{\mR})(\bar d_{\mR} \gamma^{\mu} \Gamma_d d_{\mR})  \\[0.2 cm]
&+ C_4 (\bar \nu_{\mL} \gamma^{\mu} \lambda_e \nu_{\mL})(\bar d_{\mR} \gamma^{\mu} \Gamma_d d_{\mR})+ C_4 (\bar e_{\mL} \gamma^{\mu} \lambda_e e_{\mL})(\bar d_{\mR} \gamma^{\mu} \Gamma_d d_{\mR}) \\[0.2 cm]
&  + C_6 (\bar u_{\mL} \gamma^{\mu} \lambda_u u_{\mL})(\bar e_{\mR} \gamma_{\mu} \Gamma_e e_{\mR}) + C_6 (\bar d_{\mL} \gamma^{\mu} \lambda_d d_{\mL})(\bar e_{\mR} \gamma_{\mu} \Gamma_e e_{\mR})\Big] \, . \label{eq:NP_Lambda_mass_basis}
\end{aligned}
\end{align}
From this expression we can read the independent parameters of our setup, namely the five Wilson coefficients $C_i$ and the matrices $\lambda^e$, $\lambda^d$, $\Gamma^e$ and  $\Gamma^d$. 

\begin{table}
[h]
\renewcommand{\arraystretch}{1.1}
\centering
\begin{tabular}{|c|c|c|c|}
\hline 
\multicolumn{2}{|c|}{Leptonic operators} & 
\multicolumn{2}{c|}{Semileptonic operators}\\
\hline \hline
$[Q_{\ell \ell}]_{prst}$& $(\bar \ell^{'}_{p\mL} \gamma_{\mu} \ell^{'}_{r\mL})(\bar \ell^{'}_{s\mR} \gamma^{\mu} \ell^{'}_{t\mR})$ & $[Q_{\ell q}^{\mysmall (1)}]_{prst}$ &$ (\bar \ell^{'}_{p\mL} \gamma_{\mu} \ell^{'}_{rL})(\bar q^{'}_{s\mL} \gamma^{\mu} q^{'}_{t\mL} )$\\
$[Q_{\ell e}]_{prst}$& $(\bar \ell^{'}_{p\mL} \gamma^{\mu} \ell^{'}_{r\mL})(\bar e^{'}_{s\mR} \gamma_{\mu} e^{'}_{t\mR})$ & $[Q_{\ell q}^{\mysmall (3)}]_{prst}$& $(\bar \ell^{'}_{p\mL} \gamma_{\mu} \tau^a \ell^{'}_{r\mL})(\bar q^{'}_{s\mL} \gamma^{\mu} \tau^a q^{'}_{t\mL} )$ \\
$[Q_{e e}]_{prst}$& $(\bar e'_{p\mR} \gamma_{\mu} e'_{r\mR})(\bar e'_{s\mR} \gamma^{\mu} e'_{t\mR})$ & $[Q_{\ell u}]_{prst}$& $(\bar \ell'_{p\mL} \gamma_{\mu} \ell'_{r\mL})(\bar u'_{s\mR} \gamma^{\mu} u'_{t\mR} )$ \\
 && $[Q_{\ell d}]_{prst}$& $(\bar \ell^{'}_{p\mL} \gamma_{\mu} \ell^{'}_{r\mL})(\bar d'_{s\mR} \gamma^{\mu} d'_{t\mR} )$ \\
 &&$[Q_{qe}]_{prst}$& $(\bar q'_{p\mL} \gamma_{\mu} q'_{r\mL})(\bar e'_{s\mR} \gamma^{\mu} e'_{t\mR} )$ \\
 &&$[Q_{eu}]_{prst}$& $(\bar e'_{p\mR} \gamma_{\mu} e'_{r\mR} )(\bar u'_{s\mR} \gamma^{\mu} u'_{t\mR} )$ \\
 & &$[Q_{ed}]_{prst}$& $(\bar e'_{p\mR} \gamma_{\mu} e'_{r\mR} )(\bar d'_{s\mR} \gamma^{\mu} d'_{t\mR} )$\\[0.1 cm]
\hline
\multicolumn{2}{|c|}{Vector operators} & 
\multicolumn{2}{c|}{Hadronic operators}\\
\hline \hline
$[Q_{H \ell}^{\mysmall (1)}]_{pr}$& $(\phi^\dagger i \overleftrightarrow{D_{\mu}} \phi)(\bar \ell^{\prime}_{p\mL} \gamma^{\mu} \ell^{\prime}_{r\mL}) $&$[Q_{q q}^{\mysmall (1)}]_{prst}$& $(\bar q'_{p\mL} \gamma_{\mu} q'_{r\mL})(\bar q'_{s\mL} \gamma^{\mu} q'_{t\mL})$ \\
$[Q_{H \ell}^{\mysmall (3)}]_{p\mR}$& $(\phi^\dagger i \overleftrightarrow{D_{\mu}^a} \phi)(\bar \ell^{\prime}_{p\mL} \gamma^{\mu} \tau^a \ell^{\prime}_{r\mL})$&$[Q_{q q}^{\mysmall (3)}]_{prst}$& $(\bar q'_{p\mL} \gamma_{\mu} \tau^a q'_{r\mL})(\bar q'_{s\mL} \gamma^{\mu} \tau^a q'_{t\mL})$\\
$[Q_{H q}^{\mysmall (1)}]_{p\mR}$& $(\phi^\dagger i \overleftrightarrow{D_{\mu}} \phi)(\bar q'_{p\mL} \gamma^{\mu} q'_{r\mL}) $&$[Q_{q u}^{\mysmall (1)}]_{prst}$& $(\bar q'_{p\mL} \gamma_{\mu} q'_{r\mL})(\bar u'_{s\mR} \gamma^{\mu}  u'_{t\mR})$\\
$[Q_{H q}^{\mysmall (3)}]_{p\mR}$& $(\phi^\dagger i \overleftrightarrow{D_{\mu}^a} \phi)(\bar q'_{p\mL} \gamma^{\mu} \tau^a q'_{r\mL})$&$[Q_{q d}^{\mysmall (1)}]_{prst}$& $(\bar q'_{p\mL} \gamma_{\mu}  q'_{r\mL})(\bar d'_{s\mR} \gamma^{\mu}  d'_{t\mR})$\\
$[Q_{H e}]_{p\mR}$& $(\phi^\dagger i \overleftrightarrow{D_{\mu}} \phi)(\bar e'_{p\mR} \gamma_{\mu} e'_{r\mR}) $&$[Q_{d d}]_{prst}$& $(\bar d'_{p\mR} \gamma_{\mu}  d'_{r\mR})(\bar d'_{s\mR} \gamma^{\mu}  d'_{t\mR})$\\
$[Q_{H d}]_{p\mR}$& $(\phi^\dagger i \overleftrightarrow{D_{\mu}} \phi)(\bar d'_{p\mR} \gamma^{\mu} d'_{r\mR}) $&$[Q_{u d}^{\mysmall (1)}]_{prst}$& $(\bar u'_{p\mR} \gamma_{\mu}  u'_{r\mR})(\bar d'_{s\mR} \gamma^{\mu}  d'_{t\mR})$ \\[0.1 cm]
\hline 
\end{tabular}
\caption{$SU(2)_{\mysmall L} \times U(1)_{\rm \mysmall Y}$ invariant operators involved in the renormalization group evolution of $\cL^{\mysmall 0}_{\rm \mysmall NP}$ from $\Lambda$ to the EW scale. We adopt the same notation as in \cite{Grzadkowski:2010es}.}
\label{tab:rge_flow}

\end{table} 

Following the same steps of ref. \cite{Feruglio:2016gvd,Feruglio:2017rjo}, we include RGE electroweak effects in leading logarithmic approximation. The operators involved in the running from $\Lambda$ to the EW scale are displayed in table \ref{tab:rge_flow}. We find that the effective Lagrangian at the scale $m_{\rm \mysmall EW}<\mu<\Lambda$ is given by ${\cal L} =   {\cal L}_{\rm \mysmall SM} + {\cal L}_{\rm \mysmall NP}^{\mysmall 0} + {\cal L}_{\rm eff}$, where $ {\cal L}_{\rm eff} $ describes the contribution induced by RGE and can be written as
\begin{align}
\begin{aligned}
 {\cal L}_{\rm eff} =  \delta{\cal L}_{\rm \mysmall SL} + \delta{\cal L}_{\rm \mysmall L} + \delta{\cal L}_{\rm \mysmall V} + \delta{\cal L}_{\rm \mysmall H}  \, .
 \label{eq:effective_Lagrangian}
 \end{aligned}
\end{align}
Explicitly we have 
\begin{align}
\begin{aligned}
\label{eq:semileptonic_Lagrangian}
\delta{\cal L}_{\rm \mysmall SL} = \frac{L}{16 \pi^2 \Lambda^2} &\left\lbrace (g_1^2 C_1 - 9 g_2^2 C_3)\eftQF{\ell q}{(1)}{3333}  -\frac{2}{9}g_1^2(C_1-C_4)\eftQF{\ell q}{(1)}{33ss} \right.  \\ 
&-\frac{2}{3}g_1^2 (C_1+C_6)\eftQF{\ell q}{(1)}{ss33}  -\frac{1}{2} C_1 \left([Y_u^{\dagger}Y_u]_{s3} \delta_{3t} + \delta_{s3} [Y_u^{\dagger}Y_u]_{3t} \right)\eftQF{\ell q}{(1)}{33st} \\ 
&+ \left(-3g_2^2 C_1 + C_3(6 g_2^2 + g_1^2)\right)\eftQF{\ell q}{(3)}{3333}   -2 g_2^2 C_3 \eftQF{\ell q}{(3)}{33ss} 
 -\frac{2}{3}g_2^2 C_3 \eftQF{\ell q}{(3)}{ss33}    \\ 
&  -\frac{1}{2} C_3 \left([Y_u^{\dagger}Y_u]_{s3} \delta_{3t} + \delta_{s3} [Y_u^{\dagger}Y_u]_{3t} \right) \eftQF{\ell q}{(3)}{33st} -\frac{8}{9} g_1^2 (C_1 - C_4) \eftQF{\ell u}{}{33ss}   \\ 
& +2[Y_u]_{s3} [Y_u^{\dagger}]_{3t} C_1 \eftQF{\ell u}{}{33st} +2 g_1^2 C_4 \eftQF{\ell d}{}{3333} +\frac{4}{9}g_1^2 (C_1 - C_4)\eftQF{\ell d}{}{33ss} \\
& -\frac{2}{3}g_1^2(C_4 + C_5)\eftQF{\ell d}{}{ss33} -2 g_1^2 C_6 \eftQF{qe}{}{3333} -\frac{4}{3}g_1^2 (C_1 + C_6) \eftQF{qe}{}{33ss}   \\
&  -\frac{1}{2} C_6 \left([Y_u^{\dagger}Y_u]_{s3} \delta_{3t} + \delta_{s3} [Y_u^{\dagger}Y_u]_{3t} \right)\eftQF{qe}{}{st33} + \frac{8}{9}g_1^2 (C_5-C_6)\eftQF{eu}{}{33ss} \\
& + 2[Y_u]_{s3}[Y_u^{\dagger}]_{3t}C_6 \eftQF{eu}{}{33st}-4 g_1^2 C_5 \eftQF{ed}{}{3333}+ \frac{2}{9}g_1^2 (C_5-C_6)\eftQF{qe}{}{33ss}\\
&\left. -\frac{4}{9}g_1^2 (C_5-C_6)\eftQF{ed}{}{33ss}  -\frac{4}{3}g_1^2(C_4 + C_5)\eftQF{ed}{}{ss33} \right\rbrace \, ,
\end{aligned}
\end{align}
\begin{align}
\begin{aligned}
\delta{\cal L}_{\rm \mysmall L} = \frac{L}{16 \pi^2 \Lambda^2}&\left\lbrace \left(\frac{2}{3}g_1^2 (C_1 - C_4) + 2 g_2^2 C_3\right)\eftQF{\ell\ell}{}{33ss} -4 g_2^2 C_3 \eftQF{\ell\ell}{}{3ss3} \right.  \\
& \left.+ \frac{4}{3} g_1^2(C_1 - C_4) \eftQF{\ell e}{}{33ss} -\frac{2}{3}g_1^2 (C_5-C_6) \eftQF{\ell e}{}{ss33} \right\rbrace \, ,
\label{eq:leptonic_Lagrangian}
\end{aligned}
\end{align}
\begin{align}
\begin{aligned}
\delta{\cal L}_{\rm \mysmall V} = \frac{L}{16 \pi^2 \Lambda^2} &\left\lbrace  \left(-6 C_1 \lambda^u_{33} y_t^2 - \frac{2}{3} g_1^2 (C_1 - C_4)\right) \eftQF{H \ell}{(1)}{33} \right.  \\
&  + \left(6 C_3 \lambda^u_{33} y_t^2 - 2 g_1^2 (C_1 - C_4)\right) \eftQF{H \ell}{(3)}{33} + \frac{2}{3}g_1^2 (C_1 + C_6)\eftQF{H q}{(1)}{33} \\
& -\frac{2}{3}g_2^2 C_3 \eftQF{H q}{(3)}{33} + \left( \frac{2}{3}g_1^2 (C_5-C_6) - 6 C_6 \lambda^u_{33} y_t^2 \right) \eftQF{He}{}{33} \\
&\left.+ \frac{2}{3}g_1^2 (C_4 + C_5)\eftQF{Hd}{}{33}\right\rbrace \, ,
\label{eq:vector_Lagrangian}
\end{aligned}
\end{align}
\begin{align}
\begin{aligned}
\delta{\cal L}_{\rm \mysmall H} =\frac{L}{16 \pi^2 \Lambda^2} & \left\lbrace\frac{2}{9}g_1^2 (C_1 + C_6)\eftQF{qq}{(1)}{33ss} -\frac{2}{3}g_2^2 C_3 \eftQF{qq}{(3)}{33ss} + \frac{8}{9}g_1^2( C_1 + C_6)\eftQF{qu}{(1)}{33ss} \right.  \\
& \left. -\frac{4}{9}g_1^2 (C_1 + C_6)\eftQF{qd}{(1)}{33ss} + \frac{2}{9}g_1^2(C_4 + C_5)\eftQF{qu}{(1)}{ss33} -\frac{4}{9}g_1^2(C_4 + C_5)\eftQF{dd}{}{33ss} \right\rbrace \, .
\label{eq:hadronic_Lagrangian} 
\end{aligned}
\end{align}
where $L = \log \frac{\Lambda}{\mu}$, the sum over repeated flavour indices is understood and the results are expressed in the interaction basis.
In the above expressions, we have sistematically included both gauge and top yukawa interactions, exploiting the results of~\cite{rge_top,rge_gauge}\footnote{Notice that QCD interactions do not renormalise the quark currents $V\pm A$ analysed here.}. 
Instead, we have neglected down-quark and leptons yukawas since their effects are very small.

%\red{Comment on gauge and Yukawa interaction. Only the top yukawa is non-vanishing.}\\

After the breaking of the electroweak symmetry, $\delta{\cal L}_{\rm \mysmall V}$ induces modifications of the $W$ and $Z$ couplings to fermions. The full $Z$ and $W$ Lagrangian reads:
\be
\cL_{\rm \mysmall Z,W}= -\frac{g_2}{c_{\rm \mysmall W}} Z_{\mu} J^{\mu \mysmall 0} -\frac{g_2}{\sqrt{2}} \left(W_{\mu}^+ J^{\mu, -} + \rm h.c. \right) \, ,
\label{eq:ZWfull}
\ee
where 
\noindent 
\begin{align}
& J^{\mu, \mysmall 0} =  \sum_f  \left[ (g_{ \mL, \rm \mysmall SM}^f +\Delta g_{\mL}^f)_{ij} \bar f_{i\mL} \gamma^{\mu} f_{j\mL} + (g_{ \mR, \rm \mysmall SM}^f+\Delta g_{\mR}^f)_{ij} \bar f_{i\mR} \gamma^{\mu} f_{j\mR}\right] \\
& J^{\mu, -}  = (g^{\ell}_{\rm \mysmall SM}+\Delta g^{\ell})_{ij} \bar \nu_{i \mL} \gamma^{\mu} e_{j \mL} + (g^{q}_{\rm \mysmall SM}+\Delta g^{q})_{ij} \bar u_{i\mL} \gamma^{\mu} d_{j \mL} \, ,
\label{WZinteractions}
\end{align}
and $c_{\rm \mysmall W}=\cos \theta_{\rm \mysmall W}$. These expressions include the SM contribution
\begin{align}
\begin{aligned}
&(g_{ \mL, \rm \mysmall SM}^f)_{ij}=g_{ \mL, \rm \mysmall SM}^f \delta_{ij}=(T_3^f-q_f s_{\rm \mysmall W}^2)\delta_{ij}   \\
&(g_{\mR,\rm \mysmall SM}^f)_{ij}=g_{ \mR, \rm \mysmall SM}^f \delta_{ij}= -q_f s_{\rm \mysmall W}^2\delta_{ij}  \\
&(g^{\ell}_{\rm \mysmall SM})_{ij}=\delta_{ij} \\
&(g^{q}_{\rm \mysmall SM})_{ij}= (V_{\rm \mysmall CKM})_{ij} 
\label{eq:ZW_couplings_SM}
\end{aligned}
\end{align}
and the NP contribution, encoded in the deviations $\Delta g_{\mL,\mR}^f$ and $\Delta g^{q/\ell}$. For the $Z$ couplings we have
\begin{align}
\begin{aligned}
&(\Delta g_{\mL}^{\nu})_{ij}= \frac{v^2}{\Lambda^2} \frac{L}{16 \pi^2} \left[ \frac{g_1^2}{3}(C_1 - C_4) - g_2^2 C_3 + 3 \lambda^u_{33}y_t^2(C_1+ C_3)\right]\lambda^e_{ij}   \\
&(\Delta g_{\mL}^e)_{ij}= \frac{v^2}{\Lambda^2} \frac{L}{16 \pi^2}\left[ \frac{g_1^2}{3}(C_1 - C_4) + g_2^2 C_3 + 3 \lambda^u_{33}y_t^2(C_1-C_3)\right]\lambda^e_{ij}   \\
&(\Delta g_{\mL}^u)_{ij}= \frac{v^2}{\Lambda^2} \frac{L}{16 \pi^2}  \frac{1}{3} \left[-g_2^2 C_3 - g_1^2 (C_1+C_6)\right] \lambda^u_{ij}   \\
&(\Delta g_{\mL}^d)_{ij}= \frac{v^2}{\Lambda^2} \frac{L}{16 \pi^2}  \frac{1}{3} \left[g_2^2 C_3 - g_1^2 (C_1 + C_6)\right]\lambda^d_{ij}   \\
&(\Delta g_{\mR}^e)_{ij}= \frac{v^2}{\Lambda^2} \frac{L}{16 \pi^2} \left[ -\frac{1}{3}g_1^2 (C_5-C_6) + 3 C_6 \lambda^u_{33} y_t^2 \right]\Gamma^e_{ij}   \\
&(\Delta g_{\mR}^u)_{ij}=0 \\
&(\Delta g_{\mR}^d)_{ij}= \frac{v^2}{\Lambda^2} \frac{L}{16 \pi^2} \left[ -\frac{1}{3}g_1^2 (C_4+C_5)\right]\Gamma^d_{ij} \, , \label{eq:Z_couplings_NP}
\end{aligned}
\end{align}
while for $W$ couplings we find
\begin{align}
&(\Delta g^{\ell})_{ij}= \frac{v^2}{\Lambda^2} \frac{L}{16 \pi^2} \left[6 C_3 \lambda^u_{33} y_t^2 - 2 g_2^2 C_3 \right]\lambda^e_{ij} \nn   \\
&(\Delta g^{q})_{ij}= \frac{v^2}{\Lambda^2} \frac{L}{16 \pi^2} \left[-\frac{2}{3}g_2^2 C_3\right]\lambda^{ud}_{ij}  \, .
\label{eq:W_couplings_NP}
\end{align}
We see that RGE effects induce flavour and flavour universality violating interactions, which are absent in the SM. We have explicitly checked that the dependence
on the unphysical scale $\mu$ cancels when physical quantities are computed. For $W$ and $Z$ decays, this approximately amounts to make use of $\cL_{\rm \mysmall Z,W}$
in eq. (\ref{eq:ZWfull}) in the tree-level approximation by replacing $\mu$ with the electroweak scale.

At the scale $\mu = m_{\rm \mysmall EW}$ we match the effective Lagrangian ${\cal L}_{\rm eff}$ with a new Lagrangian ${\cal L}_{\rm eff}^{\rm \mysmall EW}$ obtained by integrating out the $W$, $Z$ bosons and the top quark.  For the vector bosons $W$ and $Z$ we work at the tree-level. Disregarding the purely hadronic contribution, we get:
\begin{align}
{\cal L}_{\rm eff}^{\rm \mysmall EW}=\frac{1}{\Lambda^2}  \sum_i C_i(m_{\rm \mysmall EW}) Q_i = \frac{1}{16 \pi^2 \Lambda^2} \log \frac{\Lambda}{m_{\rm \mysmall EW}} \sum_i \xi_i Q_i \, .
\end{align}
The operators $Q_i$ and their coefficients $\xi_i$ are listed in the tables 2,3,4 and 5.
%
%\newpage
%
\begin{table}
%[t!] 
\renewcommand{\arraystretch}{1.1}
\begin{center} 
\begin{tabular}{|c|l|} 
\hline
 $Q_i$ & $\xi_i$ \\
 \hline \hline
 $(\bar \nu_{i\mL} \gamma_{\mu} \nu_{j\mL})(\bar \nu_{k\mL} \gamma_{\mu}\nu_{n\mL})$ & $\lambda^e_{ij}\delta_{kn}\left[-6 y_t^2 \lambda^u_{33} (C_1 + C_3)\right]$\\
 \hline
 $(\bar \nu_{i\mL} \gamma_{\mu} \nu_{j\mL})(\bar e_{k\mL} \gamma_{\mu} e_{n\mL})$ & $\lambda^e_{ij}\delta_{kn}\left[\frac{4}{3}e^2 \left(C_1+3C_3-C_4\right) - 12 \left(-\frac{1}{2} + s_{\rm \mysmall W}^2\right) y_t^2 \lambda^{u}_{33} \left(C_1 + C_3\right)\right]$\\
& $+\delta_{ij}\lambda^e_{kn}\left[-6 y_t^2 \lambda^u_{33} (C_1 - C_3)\right]$\\
 \hline
$(\bar \nu_{i\mL} \gamma^{\mu} \nu_{j\mL})(\bar e_{k\mR} \gamma_{\mu} e_{n\mR})$& $\lambda^e_{ij}\delta_{kn}\left[\frac{4}{3}e^2 \left(C_1+3C_3-C_4\right) - 12 s_{\rm \mysmall W}^2y_t^2 \lambda^{u}_{33} \left(C_1 + C_3)\right)\right] $\\
 & $+\delta_{ij}\Gamma^e_{kn}\left[-6 C_6 \lambda^u_{33} y_t^2\right]$\\
\hline
$(\bar e_{i\mL} \gamma^{\mu} e_{j\mL})(\bar e_{k\mL} \gamma_{\mu} e_{n\mL})$ & $\delta_{ij}\lambda^e_{kn}\left[\frac{4}{3} e^2 \left(C_1-3C_3-C_4\right) - 12 \left(-\frac{1}{2} + s_{\rm \mysmall W}^2 \right) y_t^2 \lambda^{u}_{33} \left(C_1 -C_3\right)\right]$\\
\hline
$(\bar e_{i\mL} \gamma^{\mu} e_{j\mL})(\bar e_{k\mR} \gamma_{\mu} e_{n\mR})$& $\lambda^e_{ij}\delta_{kn}\left[\frac{4}{3}e^2 \left(C_1-3C_3-C_4\right) - 12 s_{\rm \mysmall W}^2y_t^2 \lambda^{u}_{33} \left(C_1 - C_3\right)\right] $\\
 & $+\delta_{ij}\Gamma^e_{kn}\left[-\frac{4}{3}e^2 (C_5-C_6) -12 (-\frac{1}{2} + s_{\rm \mysmall W}^2) C_6 \lambda^u_{33} y_t^2\right]$\\
\hline
$(\bar e_{i\mR} \gamma^{\mu} e_{j\mR})(\bar e_{k\mR} \gamma_{\mu} e_{n\mR})$ & $\delta_{ij}\Gamma^e_{kn}\left[-\frac{4}{3}e^2 (C_5 - C_6) - 12 s_{\rm \mysmall W}^2C_6 \lambda^u_{33} y_t^2 \right]$\\
\hline
$(\bar \nu_{i\mL} \gamma_{\mu} e_{j\mL})(\bar e_{k\mL} \gamma_{\mu}  \nu_{n\mL})$ & $\left(\lambda^e_{ij}\delta_{kn}+\delta_{ij}\lambda^e_{kn}\right)\left[-12 y_t^2 \lambda^u_{33}C_3\right]$\\
\hline
\end{tabular} 
\caption{Operators $Q_i$ and coefficients $\xi_i$ for the purely leptonic part of the effective Lagrangian ${\cal L}_{\rm eff}^{\rm \mysmall  EW}$. We set $\sin^2\theta_W\equiv s^2_{\rm \mysmall W}$. \label{tab:l}}
\end{center} 
\end{table} 
\begin{table}
%[h!] 		% semileptonic 4f, \nu
\centering
\begin{tabular}{| c | l |}
		\hline
		$Q_i$& \multicolumn{1}{c |}{$\xi_i$}\\
		\hline
		&\\[-19pt]
		\hline
		$(\bar\nu_{i L}\gamma_\mu \nu_{j L}) \, (\bar u_{k L}\gamma^\mu  u_{n L})$
					&
					$ \phantom{+} \lambda^e_{ij} \, \lambda^u_{k n}    
								\left[  (g_1^2-3g_2^2)(C_1+C_3)  \right]$ \\
					&
					$ + \lambda^e_{ij} \, \delta_{kn}
								\left[  -\frac89 e^2(C_1+3C_3-C_4)-12(\frac12 - \frac23 s_{\rm \mysmall W}^2)~ y_t^2 \lambda^u_{33}(C_1+C_3)  \right]$ \\					
					&
					$ + \lambda^e_{ij} \, (\lambda^u_{k3}\delta_{3n} + \delta_{k3}\lambda^u_{3n})
								\left[ - \frac{1}{2}y_t^2(C_1+C_3)  \right]$ \\ [3pt]
		\hline
		$(\bar\nu_{i L}\gamma_\mu \nu_{j L}) \, (\bar u_{k R}\gamma^\mu  u_{n R})$
					&
					$ \phantom{+} \lambda^e_{ij} \, \delta_{kn} 
					 			\left[   -\frac89 e^2(C_1+3C_3-C_4)+8 s_{\rm \mysmall W}^2 \, y_t^2 \lambda^u_{33}(C_1+C_3)  \right]  $ \\
					&
					$ + \lambda^e_{ij} \, \delta_{k3} \delta_{3n}    
								\left[  2 y_t^2 \lambda^u_{33}  C_1  \right]$ \\ [3pt]
		\hline
		$(\bar\nu_{i L}\gamma_\mu \nu_{j L}) \, (\bar d_{k L}\gamma^\mu  d_{n L})$
					&
					$ \phantom{+} \lambda^e_{ij} \, \lambda^d_{k n}    
								\left[  (g_1^2+3g_2^2)C_1-(g_1^2+15g_2^2)C_3  \right]$ \\
					&
					$ + \lambda^e_{ij} \, \delta_{kn}
								\left[  \frac49 e^2(C_1+3C_3-C_4)-12(-\frac12+\frac13 s_{\rm \mysmall W}^2) \, y_t^2 \lambda^u_{33}(C_1+C_3)  \right]$ \\					
					&
					$ + \lambda^e_{ij} \, ((\lambda^{ud \, \dagger})_{k3}V^{\mysmall{CKM}}_{3n} 
																		+ (V^{\mysmall{CKM}})^\dagger_{k3} \lambda^{ud}_{3n})
								\left[  -\frac{1}{2}y_t^2(C_1-C_3)  \right]$ \\ [3pt]
		\hline
		$(\bar\nu_{i L}\gamma_\mu \nu_{j L}) \, (\bar d_{k R}\gamma^\mu  d_{n R})$
					&
					$ \phantom{+} \lambda^e_{ij} \, \delta_{kn}
								\left[  \frac49 e^2(C_1+3C_3-C_4)-4 s_{\rm \mysmall W}^2 \, y_t^2 \lambda^u_{33}(C_1+C_3)  \right]$ \\ 
								&$+ \delta_{ij} \, \Gamma^d_{kn} \left[2 g_1^2 C_4\right]$\\[3pt]
		\hline	
\end{tabular}
\caption{Operators $Q_i$ and coefficients $\xi_i$ for the semileptonic part of the effective Lagrangian ${\cal L}_{\rm eff}^{\rm \mysmall  EW}$ involving neutrinos and neutral currents.
Generation indices run from 1 to 3, exception made for up-type quarks where $k,n = 1,2$. We set $\sin^2\theta_W\equiv s_{\rm \mysmall W}^2$.}
\label{tab:II}
\end{table}
\begin{table}
%[tp] 		% semileptonic 4f, e
\centering
\begin{tabular}{| c | l |}
		\hline
		$Q_i$& \multicolumn{1}{c |}{$\xi_i$}\\
		\hline
		&\\[-19pt]
		\hline
		$(\bar e_{i L}\gamma_\mu e_{j L}) \, (\bar u_{k L}\gamma^\mu  u_{n L})$
					&
					$    \phantom{+} \lambda^e_{ij} \, \lambda^u_{kn}
								\left[  (g_1^2+3g_2^2)C_1-(g_1^2+15g_2^2)C_3  \right]$ \\ 
					&
					$    + \lambda^e_{ij} \, \delta_{kn}
								\left[  -\frac89 e^2(C_1-3C_3-C_4)-12(\frac12-\frac32 s_{\rm \mysmall W}^2) y_t^2 \lambda^u_{33}(C_1-C_3)  \right]$ \\ 
					&
					$    + \delta_{ij} \, \lambda^u_{kn}
								\left[  -\frac{4}{3} e^2(C_1-C_3+C_6)  \right]$ \\ 
					&
					$    + \lambda^e_{ij} \, (\lambda^u_{k3}\delta_{3n} + \delta_{k3}\lambda^u_{3n})
								\left[  -\frac12 y_t^2(C_1-C_3)  \right]$ \\ [3pt]
		\hline
		$(\bar e_{i L}\gamma_\mu e_{j L}) \, (\bar u_{k R}\gamma^\mu  u_{n R})$
					&
					$    \phantom{+} \lambda^e_{ij} \, \delta_{kn}
								\left[   -\frac89 e^2(C_1-3C_3-C_4)+8 \, s_{\rm \mysmall W}^2 \, y_t^2 \lambda^u_{33}(C_1-C_3)  \right]$ \\
					&
					$    + \lambda^e_{ij} \, \delta_{k3} \delta_{3n}
								\left[  2 y_t^2 \lambda^u_{33} C_1  \right]$ \\ [3pt]
		\hline
		$(\bar e_{i R}\gamma_\mu e_{j R}) \, (\bar u_{k L}\gamma^\mu  u_{n L})$
		&
					$    \phantom{+} \Gamma^e_{ij} \, \lambda^u_{kn}
								\left[-2 g_1^2 C_6  \right]$ \\ 
								&
					$    + \Gamma^e_{ij} \, \delta_{kn}
								\left[  \frac89 e^2(C_5-C_6)-12(\frac12-\frac23 s_{\rm \mysmall W}^2) y_t^2 \lambda^u_{33} C_6  \right]$ \\ 
					&
					$    + \delta_{ij} \, \lambda^u_{kn}
								\left[  -\frac43 e^2(C_1-C_3+C_6)  \right]$ \\
								&
					$    + \Gamma^e_{ij} \, (\lambda^u_{k3}\delta_{3n} + \delta_{k3}\lambda^u_{3n})
								\left[ -\frac12 y_t^2 C_6  \right]$ \\ [3pt]
								\hline
		$(\bar e_{i R}\gamma_\mu e_{j R}) \, (\bar u_{k R}\gamma^\mu  u_{n R})$
		&$\phantom{+} \Gamma^e_{ij} \delta_{kn} \left[\frac89 e^2(C_5-C_6)+8 s_{\rm \mysmall W}^2 y_t^2 \lambda^u_{33} C_6  \right]$\\
		& $+ \Gamma^e_{ij} \delta_{3k}\delta_{3n} \left[2 y_t^2\lambda^u_{33} C_6\right]$\\		[3pt]
		\hline
		$(\bar e_{i L}\gamma_\mu e_{j L}) \, (\bar d_{k L}\gamma^\mu  d_{n L})$
					&
					$    \phantom{+} \lambda^e_{ij} \, \lambda^d_{kn}
								\left[  (g_1^2-3g_2^2)(C_1+C_3)  \right]$ \\ 
					&
					$    + \lambda^e_{ij} \, \delta_{kn}
								\left[  \frac49 e^2(C_1-3C_3-C_4)-12(-\frac12+\frac13 s_{\rm \mysmall W}^2)\, y_t^2 \lambda^u_{33}(C_1-C_3)  \right]$ \\ 
					&
					$    + \delta_{ij} \, \lambda^d_{kn}
								\left[  -\frac43 e^2(C_1+C_3+C_6)  \right]$ \\ 
					&
					$    + \lambda^e_{ij} \, ((\lambda^{ud \, \dagger})_{k3}V^{\mysmall{CKM}}_{3n} 
																		+ {({V^{\mysmall{CKM}}}^\dagger)_{k3}} \lambda^{ud}_{3n})
								\left[  -\frac12 \, y_t^2(C_1+C_3)  \right]$ \\ [3pt]
		\hline
		$(\bar e_{i L}\gamma_\mu e_{j L}) \, (\bar d_{k R}\gamma^\mu  d_{n R})$
		&
					$    \phantom{+} \lambda^e_{ij} \, \Gamma^d_{kn}
		\left[2 g_1^2 C_4 \right]$ \\[3pt]
					&
					$    + \lambda^e_{ij} \, \delta_{kn}
								\left[  \frac49 e^2(C_1-3C_3-C_4)-4 \, s_{\rm \mysmall W}^2\, y_t^2 \lambda^u_{33}(C_1-C_3)  \right]$ \\ 
								&
					$    + \delta_{ij} \, \Gamma^d_{kn}
					\left[ -\frac43 e^2(C_4+C_5)\right]$\\ [3pt]
		\hline
		$(\bar e_{i R}\gamma_\mu e_{j R}) \, (\bar d_{k L}\gamma^\mu  d_{n L})$
		&
					$    \phantom{+} \Gamma^e_{ij} \, \lambda^d_{kn}
		\left[-2 g_1^2 C_6 \right]$ \\
		&
					$    + \Gamma^e_{ij} \, \delta_{kn}
								\left[ - \frac49 e^2(C_5-C_6)-12 \, (-\frac12+\frac13 s_{\rm \mysmall W}^2)\, y_t^2 \lambda^u_{33}C_6  \right]$ \\ 
					&
					$    + \delta_{ij} \, \lambda^d_{kn}
								\left[  -\frac43 e^2(C_1+C_3+C_6)  \right]$ \\
								&
					$    + \Gamma^e_{ij} \, ((\lambda^{ud \, \dagger})_{k3}V^{\mysmall{CKM}}_{3n} 
																		+ {({V^{\mysmall{CKM}}}^\dagger)_{k3}} \lambda^{ud}_{3n})
								\left[  -\frac12 \, y_t^2 C_6  \right]$ \\ [3pt]
								\hline
		$(\bar e_{i R}\gamma_\mu e_{j R}) \, (\bar d_{k R}\gamma^\mu  d_{n R})$
		&
		$\phantom{+} \Gamma^e_{ij} \Gamma^d_{kn} \left[-4 g_1^2 C_5   \right]$\\
		&$+\Gamma^e_{ij} \delta_{kn}\left[-\frac49 e^2(C_5-C_6)-4 s_{\rm \mysmall W}^2y_t^2 \lambda^u_{33}C_6\right]$\\
		&$+\delta_{ij} \Gamma^d_{kn}\left[-\frac43 e^2(C_4+C_5\right]$\\
[3pt]
\hline
\end{tabular}
\caption{Operators $Q_i$ and coefficients $\xi_i$ for the semileptonic part of the effective Lagrangian ${\cal L}_{\rm eff}^{\rm \mysmall EW}$ involving charged leptons and neutral currents.
Generation indices run from 1 to 3, exception made for up-type quarks where $k,n = 1,2$. We set $\sin^2\theta_W\equiv s_{\rm \mysmall W}^2$.}
\label{tab:lII}
\end{table}

\begin{table}
%[tp] 		% semileptonic 4f, CC
\centering
\begin{tabular}{| c | l |}
		\hline
		$Q_i$& \multicolumn{1}{c |}{$\xi_i$}\\
		\hline
		&\\[-19pt]
		\hline
		$(\bar e_{i L}\gamma_\mu \nu_{j L}) (\bar u_{k L}\gamma^\mu  d_{n L})$
					&
					$   \phantom{+} \lambda^e_{ij} \, \lambda^{ud}_{kn}
								\left[  -6 g_2^2 C_1+2(6g_2^2+g_1^2)C_3  \right]$ \\
					&
					$   + \lambda^e_{ij} \, V^{\mysmall{CKM}}_{kn}
								\left[  -12 \, y_t^2 \lambda^u_{33} C_3  \right]$ \\ 
					&
					$   + \lambda^e_{ij} \, ( \lambda^u_{k3} V^{\mysmall{CKM}}_{3n} +
																			\delta_{k3} \lambda^{ud}_{3n} )
								\left[  - y_t^2 C_3  \right]$ \\ [3pt]
\hline
\end{tabular}
\caption{Operators $Q_i$ and coefficients $\xi_i$ for the semileptonic part of the effective Lagrangian ${\cal L}_{\rm eff}^{\rm \mysmall EW}$ involving charged currents. 
For up-type quarks the indices run from 1 to 2. The $\xi_i$ coefficient for the Hermitian conjugate operator can be easily derived. \label{tab:4fCC}}
\end{table}

%\newpage

Below the electroweak scale only the residual electromagnetic gauge symmetry is relevant to our discussion, and the effective theory consists of a combination of 
$U(1)_{\rm em}$-invariant operators whose Wilson coefficients run under the effect of QED interactions only. By lowering the scale $\mu$ we first cross the bottom quark
mass threshold, then the charm one.  When crossing a threshold we integrate out the corresponding quark and match the theory to a new one.
At the scale $\mu\approx 1$ GeV we get the following result for the effective Lagrangian ${\cal L}^{\rm \mysmall QED}_{\rm eff}$:
\begin{align}
\begin{aligned}
{\cal L}^{\rm \mysmall QED}_{\rm eff} &= \frac{1}{\Lambda^2} \sum_i {\cal C}_i (m_{\rm \mysmall EW}) Q_i  +  \frac{1}{\Lambda^2} \sum_i \delta {\cal C}_i (\mu)  Q_i^{\rm em}  \\
&=  \frac{1}{16 \pi^2 \Lambda^2}\log \frac{\Lambda}{m_{\rm \mysmall EW}} \sum_i \xi_i Q_i  + \frac{1}{16 \pi^2 \Lambda^2}\log \frac{m_{\rm \mysmall EW}}{\mu} \sum_i \delta\xi_i Q_i^{\rm em} \,  ,
\label{eq:effective_Lagrangian_GeV}
\end{aligned}
\end{align}
where the $U(1)_{\rm em}$-invariant operators $Q_i^{\rm em}$ and their coefficients $\delta\xi_i$ are collected in tables 6, 7, 8 and 9.

\vspace{1.cm}\noindent

%\newpage
\begin{table}
%[t] 		% QED leptonic 4f
\centering
\begin{tabular}{| c | l |}
		\hline
		\multicolumn{1}{| c |}{$Q_i^{\rm em}$} & \multicolumn{1}{c |}{$\delta\xi_i$}\\
		\hline
		&\\[-19pt]
		\hline
		$(\bar\nu_{i L}\gamma_\mu \nu_{j L})~(\bar \nu_{k L}\gamma^\mu  \nu_{n L})$
					&
					$ \phantom{+} 0$ \\ [3pt]
		\hline
		$(\bar\nu_{i L}\gamma_\mu \nu_{j L})~(\bar e_{k}\gamma^\mu  e_{n})$
					&
					$ \phantom{+} \lambda^e_{ij} \, \delta_{kn}
								\cdot  \frac43 e^2
								\left[  (C_1+3C_3-C_4) - 2(C_1+C_3) ({\lambda}^u_{33}
																									+\hat{\lambda}^u_{22}\log\frac{m_c}{\mu})\right.$ \\
								&
										$ \phantom{+ \lambda^e_{ij} \, \delta_{kn} \cdot  \frac43 e^2 [}
										\left. +\left((C_1-C_3)\hat{\lambda}^d_{33}+C_4 \hat\Gamma^d_{33}\right)\log\frac{m_b}{\mu}  \right]$ \\ [5pt]
		\hline
		$(\bar e_{i L}\gamma_\mu e_{j L})~(\bar e_{k}\gamma^\mu  e_{n})$
					&
					$ \phantom{+} \lambda^e_{ij} \, \delta_{kn}
								\cdot  \frac43 e^2
								\left[  (C_1-3C_3-C_4) - 2(C_1-C_3) ({\lambda}^u_{33}
																									+\hat{\lambda}^u_{22}\log\frac{m_c}{\mu})\right.$ \\
								&
										$ \phantom{+ \lambda^e_{ij} \, \delta_{kn} \cdot  \frac43 e^2 [}
										\left. +\left((C_1+C_3)\hat{\lambda}^d_{33}+C_4 \hat\Gamma^d_{33}\right)\log\frac{m_b}{\mu}  \right]$ \\
										\hline
		$(\bar e_{i R}\gamma_\mu e_{j R})~(\bar e_{k}\gamma^\mu  e_{n})$
					&
					$ \phantom{+} \Gamma^e_{ij} \, \delta_{kn}\cdot 
					\frac43 e^2\left[ (C_6-C_5) -2 C_6 ({\lambda}^u_{33}+\hat{\lambda}^u_{22}\log\frac{m_c}{\mu}) \right.$\\
&$\phantom{+ \lambda^e_{ij} \, \delta_{kn} \cdot  \frac43 e^2 [}					
					\left. +\left(C_6\hat{\lambda}^d_{33}+C_5 \hat\Gamma^d_{33}\right)\log\frac{m_b}{\mu} \right]$\\
										 [5pt]
\hline
\end{tabular}
\caption{Operators $Q_i^{\rm em}$ and coefficients $\delta\xi_i$ for the purely leptonic part of the effective Lagrangian $\delta{\cal L}_{\rm eff}^{\rm \mysmall QED}$. We set $\hat{\lambda}^{u,d}_{ii}=\lambda^{u,d}_{ii}/\log\frac{m_{\rm \mysmall EW}}{\mu}$.
\label{tab:QEDlep}}
\end{table}

\begin{table}
%[t] 		% QED semileptonic 4f, \nu
\centering
\begin{tabular}{| c | l |}
		\hline
		$Q_i^{em}$& \multicolumn{1}{c |}{$\delta\xi_i$}\\
		\hline
		&\\[-19pt]
		\hline
		$(\bar\nu_{i L}\gamma_\mu \nu_{j L}) \, (\bar u_{k}\gamma^\mu  u_{n})$
					&
					$ \phantom{+}   \lambda^e_{ij} \, \delta_{kn}  \left(-\frac89 e^2 \right)
								\left[   (C_1+3C_3-C_4)-2(C_1+C_3)({\lambda}^u_{33}+\hat{\lambda}^u_{22}\log\frac{m_c}{\mu})  \right.$ \\
								 &
								 $ \phantom{+ \lambda^e_{ij} \, \delta_{kn}  \left(-\frac89 e^2 \right) [ }
								 \left. +\left((C_1-C_3)\hat{\lambda}^d_{33}+C_4\hat\Gamma^d_{33}\right)\log\frac{m_b}{\mu} \right]$ \\ [5pt]
		\hline
		$(\bar\nu_{i L}\gamma_\mu \nu_{j L}) \, (\bar d_{k}\gamma^\mu  d_{n})$
					&
					$ \phantom{+}   \lambda^e_{ij} \, \delta_{kn}  \cdot \frac49 e^2 
								\left[   (C_1+3C_3-C_4)-2(C_1+C_3)({\lambda}^u_{33}+\hat{\lambda}^u_{22}\log\frac{m_c}{\mu})  \right.$ \\
								 &
								 $ \phantom{+ \lambda^e_{ij} \, \delta_{kn}  \cdot \frac49 e^2 [ }
								 \left. +\left((C_1-C_3)\hat{\lambda}^d_{33}+C_4\hat\Gamma^d_{33}\right)\log\frac{m_b}{\mu} \right]$ \\ [5pt]
\hline
\end{tabular}
\caption{Operators $Q_i^{\rm em}$ and coefficients $\delta\xi_i$ for the semileptonic part of the effective Lagrangian $\delta{\cal L}_{\rm eff}^{\rm \mysmall QED}$ involving neutrinos and neutral currents.
For the down-type quarks generation indices run from 1 to 2,while for up-type quarks we only keep the first generation. We set $\hat{\lambda}^{u,d}_{ii}=\lambda^{u,d}_{ii}/\log\frac{m_{EW}}{\mu}$.}
\end{table}

\begin{table}
%[hp] 		% QED semileptonic 4f, e
\centering
\begin{tabular}{| c | l |}
		\hline
		$Q_i^{em}$& \multicolumn{1}{c |}{$\delta\xi_i$}\\
		\hline
		&\\[-19pt]
		\hline
		$(\bar e_{i L}\gamma_\mu e_{j L}) \, (\bar u_{k L}\gamma^\mu  u_{n L})$
					&
					$    \phantom{+} \lambda^e_{ij} \, \lambda^u_{kn} 
								\left[ 8 e^2 \,(C_1-C_3) \right]$ \\ 
								&
					$    - \lambda^e_{ij} \, \delta_{kn}  \cdot
								 \frac89 e^2\left[(C_1-3C_3-C_4)-2(C_1-C_3)({\lambda}^u_{33}
																							+\hat{\lambda}^u_{22}\log\frac{m_c}{\mu})\right.$ \\
								&
								$ \phantom{ - \lambda^e_{ij} \, \delta_{kn}  \cdot \frac89 e^2 [ }
								\left.+\left((C_1+C_3)\hat{\lambda}^d_{33}+C_4\hat\Gamma^d_{33}\right)\log\frac{m_b}{\mu}  \right]$\\
					&
					$    + \delta_{ij} \, \lambda^u_{kn}
								\left[  -\frac{4}{3} e^2(C_1-C_3+C_6)  \right]$ 
					 \\ [3pt]
		\hline
		$(\bar e_{i L}\gamma_\mu e_{j L}) \, (\bar u_{k R}\gamma^\mu  u_{n R})$
					&
					$ - \lambda^e_{ij} \, \delta_{kn}  \cdot
								 \frac89 e^2\left[(C_1-3C_3-C_4)-2(C_1-C_3)({\lambda}^u_{33}
																							+\hat{\lambda}^u_{22}\log\frac{m_c}{\mu})\right.$ \\
								&
								$ \phantom{ - \lambda^e_{ij} \, \delta_{kn}  \cdot \frac89 e^2 [ }
								\left.+\left((C_1+C_3)\hat{\lambda}^d_{33}+C_4\hat\Gamma^d_{33}\right)\log\frac{m_b}{\mu}  \right]$ \\ [3pt]
		\hline
		$(\bar e_{i R}\gamma_\mu e_{j R}) \, (\bar u_{k L}\gamma^\mu  u_{n L})$
				&
								$ \phantom{+} \Gamma^e_{ij} \, \lambda^u_{kn}\left[-8 e^2 C_6\right]$\\	
		&
					$ +\Gamma^e_{ij} \, \delta_{kn}\cdot 
					\frac89 e^2\left[ (C_5-C_6)+2 C_6 ({\lambda}^u_{33}+\hat{\lambda}^u_{22}\log\frac{m_c}{\mu}) \right.$\\
&$\phantom{+ \lambda^e_{ij} \, \delta_{kn} \cdot  \frac43 e^2 [}					
					\left. -\left(C_6\hat{\lambda}^d_{33}+C_5 \hat\Gamma^d_{33}\right)\log\frac{m_b}{\mu} \right]$\\
					&
					$    + \delta_{ij} \, \lambda^u_{kn}
								\left[  -\frac{4}{3} e^2(C_1-C_3+C_6)  \right]$ \\ 
													[3pt]
\hline
		$(\bar e_{i R}\gamma_\mu e_{j R}) \, (\bar u_{k R}\gamma^\mu  u_{n R})$
&
					$ +\Gamma^e_{ij} \, \delta_{kn}\cdot 
					\frac89 e^2\left[ (C_5-C_6)+2 C_6 ({\lambda}^u_{33}+\hat{\lambda}^u_{22}\log\frac{m_c}{\mu}) \right.$\\
&$\phantom{+ \lambda^e_{ij} \, \delta_{kn} \cdot  \frac43 e^2 [}					
					\left. -\left(C_6\hat{\lambda}^d_{33}+C_5 \hat\Gamma^d_{33}\right)\log\frac{m_b}{\mu} \right]$\\[3pt]
		\hline
		$(\bar e_{i L}\gamma_\mu e_{j L}) \, (\bar d_{k L}\gamma^\mu  d_{n L})$
					&
					$    \phantom{+} \lambda^e_{ij} \, \lambda^d_{kn}
								\left[  -4 e^2 \,(C_1+C_3)  \right]$ \\ 
								&

					$    + \lambda^e_{ij} \, \delta_{kn} \cdot \frac49 e^2
								\left[(C_1-3C_3-C_4)-2(C_1-C_3)({\lambda}^u_{33}
																							+\hat{\lambda}^u_{22}\log\frac{m_c}{\mu})\right.$ \\
								&
								$ \phantom{ - \lambda^e_{ij} \, \delta_{kn}  \cdot \frac89 e^2 [ }
								\left.+\left((C_1+C_3)\hat{\lambda}^d_{33}+C_4\hat\Gamma^d_{33}\right)\log\frac{m_b}{\mu}  \right]$ \\
					&
					$    + \delta_{ij} \, \lambda^d_{kn}
								\left[  -\frac43 e^2(C_1+C_3+C_6)  \right]$ \\ 
					 [3pt]
		\hline
		$(\bar e_{i L}\gamma_\mu e_{j L}) \, (\bar d_{k R}\gamma^\mu  d_{n R})$
		&
		$\phantom{+} \lambda^e_{ij} \, \Gamma^d_{kn}
		\left[4 e^2 C_4\right]$\\
					&
					$    + \lambda^e_{ij} \, \delta_{kn} \cdot \frac49 e^2
								\left[(C_1-3C_3-C_4)-2(C_1-C_3)({\lambda}^u_{33}
																							+\hat{\lambda}^u_{22}\log\frac{m_c}{\mu})\right.$ \\
								&
								$ \phantom{ - \lambda^e_{ij} \, \delta_{kn}  \cdot \frac89 e^2 [ }
								\left.+\left((C_1+C_3)\hat{\lambda}^d_{33}+C_4\hat\Gamma^d_{33}\right)\log\frac{m_b}{\mu}  \right]$ \\ 
								&
		$+ \delta_{ij} \, \Gamma^d_{kn}
\cdot \left[-\frac43 e^2 (C_4+C_5)\right]$\\						[3pt]
		\hline
		$(\bar e_{i R}\gamma_\mu e_{j R}) \, (\bar d_{k L}\gamma^\mu  d_{n L})$
		&
		$\phantom{+} \Gamma^e_{ij} \, \lambda^d_{kn}
		\left[4 e^2 C_6\right]$\\
		&
		$ +\Gamma^e_{ij} \, \delta_{kn}\cdot 
					\left(-\frac49 e^2\right)\left[ (C_5-C_6)+2 C_6 ({\lambda}^u_{33}+\hat{\lambda}^u_{22}\log\frac{m_c}{\mu}) \right.$\\
&$\phantom{+ \lambda^e_{ij} \, \delta_{kn} \cdot  \frac43 e^2 [}					
					\left. -\left(C_6\hat{\lambda}^d_{33}+C_5 \hat\Gamma^d_{33}\right)\log\frac{m_b}{\mu} \right]$\\
					&
					$    + \delta_{ij} \, \lambda^d_{kn}
								\left[  -\frac43 e^2(C_1+C_3+C_6)  \right]$ \\
								 [3pt]
								 \hline
		$(\bar e_{i R}\gamma_\mu e_{j R}) \, (\bar d_{k R}\gamma^\mu  d_{n R})$
		&
		$\phantom{+} \Gamma^e_{ij} \, \Gamma^d_{kn}
		\left[-4 e^2 C_5\right]$\\
&
					$ +\Gamma^e_{ij} \, \delta_{kn}\cdot 
					\left(-\frac49 e^2\right)\left[ (C_5-C_6)+2 C_6 ({\lambda}^u_{33}+\hat{\lambda}^u_{22}\log\frac{m_c}{\mu}) \right.$\\
&$\phantom{+ \lambda^e_{ij} \, \delta_{kn} \cdot  \frac43 e^2 [}					
					\left. -\left(C_6\hat{\lambda}^d_{33}+C_5 \hat\Gamma^d_{33}\right)\log\frac{m_b}{\mu} \right]$\\
					&
		$+ \delta_{ij} \, \Gamma^d_{kn}
\cdot \left[-\frac43 e^2 (C_4+C_5)\right]$\\[3pt]
\hline
\end{tabular}
\caption{Operators $Q_i^{\rm em}$ and coefficients $\delta\xi_i$ for the semileptonic part of the effective Lagrangian $\delta{\cal L}_{\rm eff}^{\rm \mysmall QED}$ involving charged leptons and neutral currents.
For the down-type quarks generation indices run from 1 to 2, while for up-type quarks we only keep the first generation. We set  $\hat{\lambda}^{u,d}_{ii}=\lambda^{u,d}_{ii}/\log\frac{m_{\rm \mysmall EW}}{\mu}$.}
\end{table}

%%%%%%%%%%%%%%%%%%%%%%%%%%%%%%%%%%%%%%%%%%%%%%%%%%%%%%%%%%%
\section{Observables}
\label{sec_3}
%%%%%%%%%%%%%%%%%%%%%%%%%%%%%%%%%%%%%%%%%%%%%%%%%%%%%%%%%%%

This section addresses the phenomenological consequences of Lagrangian (\ref{eq:NP_Lambda}), making use of the 
RGE-improved low-energy effective field theory (EFT) derived in the previous section. 
The NP contribution to the observables is parametrised in terms of the free parameters of $\cL_{\rm \mysmall NP}^{\mysmall 0}$, 
namely the five $C_i$ and the matrices $\lambda^e$, $\lambda^d$, $\Gamma^e$ and $\Gamma^d$. 
In order to simplify our phenomenological analysis, we assume real entries in $\lambda^{e/d}$ and $\Gamma^{e/d}$,
negligible mixing with the first generation in the matrices 
$\lambda^{e/d}$ and $\Gamma^{e/d}$, $\lambda^{e/d}_{1i}=\Gamma^{e/d}_{1i}=0$ $(i=1,2,3)$
and a small mixing approximation \footnote{The largest mixing arises from $\lambda^e_{23}\approx 0.3$. In our numerical
analysis we will let $|\lambda^e_{23}|$ and $|\Gamma^e_{23}|$ vary up to $0.5$ by using complete formulae.}, implying
$$\lambda^{e/d}_{22} \approx |\lambda^{e/d}_{23}|^2 \ll \lambda^{e/d}_{33}~~~~~~~~~~~~~\lambda^{e/d}_{33} \approx 1~~~,$$
and similarly for $\Gamma^{e/d}$.
As a result, the parameters involved in our analysis are $C_1$, $C_3$, $C_4$, $C_5$, $C_6$, $\lambda^{e/d}_{23}$, $\Gamma^{e/d}_{23}$.
Beyond semileptonic B-decays, we focus on fully leptonic processes and leptonic decays of the $Z$ vector boson
as they are the only processes that compete with semileptonic B-decays in constraining our NP parameter space.
The structure of this section is as follows.  
In section 3.1, we discuss how to address both charged- and neutral-current $B$ anomalies within our framework. 
In section 3.2, we discuss the most relevant tree-level phenomenology connected with the $B$ anomalies. 
In section 3.3, we proceed to study observables in the leptonic sector receiving large contributions at loop-level.
In section 3.4, a global numerical analysis is performed in a phenomenologically relevant scenario, where NP affects 
dominantly the Wilson coefficient $\cC^{\mysmall 9}$.

%%%%%%%%%%%%%%%%%%%%%%%%%%%%%%%%%%%%%%%%%%%%%%%%%%%%%%%%%%%
\subsection{The \texorpdfstring{$B$}{B} anomalies}

The most significant measurements related to charged- and neutral-current B-anomalies are:
\begin{align}
R^{\tau/\ell}_{D^*} &=\frac{ \cB(B \to D^* \tau \overline{\nu})_{\rm exp}/\cB(B \to D^* \tau \overline{\nu})_{\rm SM} }{ \cB(B \to D^* \ell \overline{\nu} )_{\rm exp}/ \cB(B \to D^* \ell \overline{\nu} )_{\rm SM} } = 1.23 \pm 0.07~, \label{eq:RDexp}  \\
R^{\tau/\ell}_{D } &= \frac{ \cB(B \to D  \tau \overline{\nu})_{\rm exp}/\cB(B \to D  \tau \overline{\nu})_{\rm SM} }{ \cB(B \to D  \ell \overline{\nu} )_{\rm exp}/ \cB(B \to D  \ell \overline{\nu} )_{\rm SM} } =   1.34 \pm 0.17~, \label{eq:RDSexp}
\end{align}
where $\ell=e, \mu$, which follow from the HFAG averages~\cite{Amhis:2016xyh} of Babar~\cite{Lees:2013uzd}, Belle~\cite{Hirose:2016wfn}, 
and LHCb data~\cite{Aaij:2015yra}, combined with the SM predictions~\cite{Fajfer:2012vx,Aoki:2016frl}, and
\begin{align}
\RKs &=  \left. \frac{ \cB(B \to K^* \mu \bar{\mu})_{\rm exp} }{ \cB(B \to K^* e \bar{e} )_{\rm exp} } \right|_{q^2\in[1.1,6]{\rm GeV}} =  0. 685 {}^{+0.113}_{-0.069} \pm 0.047~,
\label{eq:RKSexp} \\
\RK &=  \left. \frac{ \cB(B \to K \mu \bar{\mu})_{\rm exp} }{ \cB(B \to K e \bar{e} )_{\rm exp} } \right|_{q^2\in[1,6]{\rm GeV}} =  0. 745 {}^{+0.090}_{-0.074} \pm 0.036~,
\label{eq:RKexp}
\end{align}
based on combination of LHCb data~\cite{Aaij:2017vbb} with the SM expectation $\RKss =1.00 \pm 0.01$~\cite{Bordone:2016gaq}.

We recall that $b\to s$ semileptonic transitions are conventionally described by means of the effective Lagrangian $\mathcal{L}_{\rm eff}^{\rm \mysmall NC}$\footnote{In our analysis, the inclusion of dipole operators is not necessary as they provide negligible effects.}:
\begin{equation}
\mathcal{L}_{\rm eff}^{\rm \mysmall NC} = \frac{4G_F}{\sqrt2}\,\lambda_{bs}^{t}  \Big( \mathcal{C}^{9}_{ij}\,\mathcal{O}^{9}_{ij} + \mathcal{C}^{9^{\prime}}_{ij}\,\mathcal{O}^{9^{\prime}}_{ij} + \mathcal{C}^{10}_{ij}\,\mathcal{O}^{10}_{ij} + \mathcal{C}^{10^{\prime}}_{ij}\,\mathcal{O}^{10^{\prime}}_{ij} +\mathcal{C}^{\nu}_{ij}\,\mathcal{O}^{\nu}_{ij} + \mathcal{C}^{\nu^{\prime}}_{ij}\,\mathcal{O}^{\nu^{\prime}}_{ij}\Big)\,,
\label{eq:NCeff}
\end{equation}
where $\lambda^t_{bs}=V_{tb}V_{ts}^*$ and the operators $\mathcal{O}^{i}$ are given by
\begin{align}\label{eq:opNCeff}
\begin{aligned}
\mathcal{O}^{9}_{ij}&=\frac{e^2}{(4\pi)^2}\left(\overline{s}\gamma_\mu P_Lb\right)\left(\overline{e}_i \gamma^\mu e_j\right)\,,    &\mathcal{O}^{9^\prime}_{ij} &=\frac{e^2}{(4\pi)^2} \left(\overline{s}\gamma_\mu P_Rb\right)\left(\overline{e}_i\gamma^\mu e_j\right)\,,\\[3pt] 
\mathcal{O}^{10}_{ij}&=\frac{e^2}{(4\pi)^2} \left(\overline{s}\gamma_\mu P_Lb\right)\left(\overline{e}_i\gamma^\mu\gamma_5 e_j\right)\,,   &\mathcal{O}^{10^\prime}_{ij}&=\frac{e^2}{(4\pi)^2} \left(\overline{s}\gamma_\mu P_Rb\right)\left(\overline{e}_i\gamma^\mu\gamma_5 e_j\right)\,,    \\[3pt]
\mathcal{O}^{\nu}_{ij}&=\frac{e^2}{(4\pi)^2} \left(\overline{s}\gamma_\mu P_L b\right)\left(\overline{\nu}_i\gamma^\mu (1-\gamma_5) \nu_j\right)\,,    &\mathcal{O}^{\nu^\prime}_{ij} &=  \frac{e^2}{(4\pi)^2} \left(\overline{s}\gamma_\mu P_Rb\right)\left(\overline{\nu}_i\gamma^\mu (1-\gamma_5) \nu_j\right)\,.\\[3pt] 
\end{aligned}
\end{align}
As to the charged-current transition $b \rightarrow c \ell \nu $, we address it using the effective Lagrangian $\mathcal{L}_{\rm eff}^{\rm \mysmall CC}$, defined as
\begin{equation}
\mathcal{L}_{\rm eff}^{\rm \mysmall CC} = -\frac{4G_F}{\sqrt2}\,
{{\cal C}_{\rm \mysmall L}^{cb}}^{ij}\,(\bar c_{\mL} \gamma^{\mu} b_{\mL})(\bar e_{Li} \gamma_{\mu} \nu_{\mL j})\,.\label{eq:CCeff}
\end{equation}
In our framework $B$ anomalies receive NP contributions at tree level. These contributions can be computed explicitly 
by matching the low-energy Lagrangians in eqs.~\ref{eq:NCeff}, \ref{eq:CCeff} with the NP Lagrangian $\cL_{\rm \mysmall NP}^{\mysmall 0}$ 
\footnote{Strictly speaking  $\cL^{\rm \mysmall NC}_{\rm eff}$ and $\cL^{\rm \mysmall CC}_{\rm eff}$ should be matched to the Lagrangian 
obtained by running the Wilson coefficients down to $\mu= m_{\mysmall B}$, but RGE induced terms are generally negligible 
with respect to tree-level ones. This is true unless accidental cancellations among parameters take place, which we exclude.}. 
As a result, we find 
%
%\begin{empheq}[left=\empheqlbrace]{align}
\begin{align}
\begin{aligned}
({\cal C}^{ 9}_{\rm \mysmall NP})_{ij}&= \frac{4 \pi^2}{e^2 \lambda_{bs}^t} \frac{v^2}{\Lambda^2} \lambda^d_{23} \left[(C_1 + C_3)\lambda^e_{ij}+ C_6\Gamma^e_{ij} \right]  &({\cal C}^{9^{\prime}}_{\rm \mysmall NP})_{ij}& =\frac{4 \pi^2}{e^2 \lambda_{bs}^t} \frac{v^2}{\Lambda^2} \Gamma^d_{23} \left[C_4\lambda^e_{ij} + C_5\Gamma^e_{ij} \right] \\
({\cal C}^{ 10}_{\rm \mysmall NP})_{ij}&= \frac{4 \pi^2}{e^2 \lambda_{bs}^t} \frac{v^2}{\Lambda^2} \lambda^d_{23}\left[-(C_1 + C_3)\lambda^e_{ij} + C_6\Gamma^e_{ij} \right]  &({\cal C}^{10^{\prime}}_{\rm \mysmall NP})_{ij}& =\frac{4 \pi^2}{e^2 \lambda_{bs}^t} \frac{v^2}{\Lambda^2} \Gamma^d_{23}\left[-C_4\lambda^e_{ij} + C_5\Gamma^e_{ij} \right] \\
({\cal C}^\nu_{\rm \mysmall NP})_{ij}&=\frac{4 \pi^2}{e^2 \lambda_{bs}^t} \frac{v^2}{\Lambda^2} \lambda^d_{23}  \lambda^e_{ij}(C_1 - C_3) &({\cal C}^{\nu^{\prime}}_{\rm \mysmall NP})_{ij}&=\frac{4 \pi^2}{e^2 \lambda_{bs}^t} \frac{v^2}{\Lambda^2} \Gamma^d_{23} \lambda^e_{ij} C_4  \\ 
({\cal C}^{cb}_{\mL, \rm \mysmall NP})_{ij}&= - \frac{v^2}{\Lambda^2} \lambda^{ud}_{23}  \lambda^{e}_{ij}  C_3 \,  ,
\label{eq:tree_matching}
\end{aligned}
\end{align}
%\end{empheq}
%
where subleading RGE terms have been neglected. 

We remind that NP should contribute dominantly to charged-current transitions compared to the neutral-current ones, since in the SM 
the former arise at the tree-level while the latter at one-loop. This can be achieved in our framework by assuming a hierarchy between $\lambda^d_{33}\lambda^e_{33}$ and $\lambda^d_{23}\lambda^e_{22}$, which control $B \to D^{(*)} \tau \nu$ and $B\to K\mu^+\mu^-$,
respectively.

%
%%%%%%%%%%%%%%%%%%%%%%%%%%%%%%%%%%%%%%%%%%%%%%%%%%%%%%%%%%%
\subsubsection{\texorpdfstring{$B\to K\ell\bar\ell$}{B to Kll}}
%%%%%%%%%%%%%%%%%%%%%%%%%%%%%%%%%%%%%%%%%%%%%%%%%%%%%%%%%%%
%
Keeping only linear terms in NP contributions, $R_{K^(*)}^{\mu/e}$ can be written in our framework  as~\cite{flavio} 
\begin{align}
R^{\mu/e}_{K}  \simeq&\; 1  + 0.24 [({\cal C}^{9}_{\rm \mysmall NP})_{\mu \mu} + ({\cal C}^{9^{\prime}}_{\rm \mysmall NP})_{\mu \mu}] 
- 0.26 [({\cal C}^{10}_{\rm \mysmall NP})_{\mu \mu} + ({\cal C}^{10^{\prime}}_{\rm \mysmall NP})_{\mu \mu}] 
\,,  
\nonumber \\
R^{\mu/e}_{K^\ast}   \simeq&\; 1 + 0.19 [({\cal C}^{9}_{\rm \mysmall NP})_{\mu \mu} - ({\cal C}^{9^{\prime}}_{\rm \mysmall NP})_{\mu \mu}] 
- 0.29 ({\cal C}^{10}_{\rm \mysmall NP})_{\mu \mu}  + 0.22 ({\cal C}^{10^{\prime}}_{\rm \mysmall NP})_{\mu \mu} \,,
\label{eq:RK_RKstar_NP}
\end{align}
where $({\cal C}^{ 9}_{\rm \mysmall NP})_{ee}$, $({\cal C}^{ 9^{\prime}}_{\rm \mysmall NP})_{ee}$, $({\cal C}^{10}_{\rm \mysmall NP})_{ee}$ and $({\cal C}^{10^{\prime}}_{\rm \mysmall NP})_{ee}$ can be neglected because $\lambda^e_{11} = 0$. 
Remembering that ${\cal C}^{9}_{\rm \mysmall SM} \approx - {\cal C}^{10}_{\rm \mysmall SM} \approx 4.2$ \cite{Blake:2016olu}, we find the numerical expressions 
\begin{align}
R_K^{\mu/e}& \approx 1- \frac{0.30}{\Lambda^2 ({\rm TeV}^2)} \frac{\lambda^e_{22}}{10^{-3}} \left[(C_1 + C_3)\lambda^d_{23} +  C_4\Gamma^d_{23}\right] + \cdots\, , 
\nonumber\\
R_{K^*}^{\mu/e}& \approx 
1- \frac{0.29}{\Lambda^2 ({\rm TeV}^2)} \frac{1}{10^{-3}} \left[(C_1 + C_3)\lambda^e_{22} \lambda^d_{23} -  
0.9\, C_4 \lambda^e_{22} \Gamma^d_{23} - 0.2\, C_6\lambda^d_{23} \Gamma^e_{22}\right] + \cdots \, ,
\label{eq:RK_RKstar_NP_num}
\end{align}
and dots stand for smaller contributions.
From (\ref{eq:RK_RKstar_NP_num}) and the current experimental results (\ref{eq:RKSexp}) and (\ref{eq:RKexp}), we argue that 
a simultaneous explanation of $\RK$ and $R_{K^*}^{\mu/e}$ requires the condition $|(C_1 + C_3)\lambda^d_{23}| \gg  |C_4\Gamma^d_{23}|$
and $(C_1 + C_3)\lambda^d_{23}\lambda^e_{23} \approx \mathcal{O}(10^{-3})$.

.

%%%%%%%%%%%%%%%%%%%%%%%%%%%%%%%%%%%%%%%%
\subsubsection{\texorpdfstring{$B \to D^{\mysmall (*)} \ell \nu$}{B to Dlog}}
%%%%%%%%%%%%%%%%%%%%%%%%%%%%%%%%%%%%%%%%

LFUV in the charged-current process $B \to D^{\mysmall (*)} \ell \nu$ is encoded in the observable $R^{\tau/\ell}_{D^{\mysmall (*)}}$,
which can be expressed as
\begin{equation}
R^{\tau/\ell}_{D^{\mysmall (*)}} = \frac{\sum_{j} |({\cal C}^{cb}_{\mL})_{3j}|^2}{\sum_{j} |({\cal C}^{cb}_{\mL})_{\ell j}|^2}\,.
\end{equation}
Keeping only linear NP contributions and neglecting $\lambda^e_{11}$ and $\lambda^e_{22}$ with respect to $\lambda^e_{33}$, we find
\begin{align}
R^{\tau/\ell}_{D^{\mysmall (*)}}& \approx 1- 2 \frac{v^2}{\Lambda^2} \frac{\lambda^{ud}_{23}}{V_{cb}} C_3 \lambda^e_{33} \,. 
\end{align}
Then, using the relation $\lambda^{ud}=V_{\rm \mysmall CKM} \lambda^d$, we end up with the following expression
\begin{align}
R^{\tau/\ell}_{D^{\mysmall (*)}}& \approx 1- 0.12 \frac{C_3}{\Lambda^2 ({\rm TeV}^2)} \lambda^e_{33} \left(\frac{V_{cs}}{V_{cb}} \lambda^d_{23} + \lambda^d_{33} \right) \, .
\end{align}
As a result, in order to accommodate the $R^{\tau/\ell}_{D^{(*)}}$ anomaly, 
we need $C_3<0$ and $C_3 \sim \mathcal{O}(1)$, for $\Lambda=1$ TeV.

%
%%%%%%%%%%%%%%%%%%%%%%%%%%%%%%%%%%%%%%%%
\subsection{Tree-level semileptonic phenomenology}
Our framework predicts a set of deviations in leptonic and semileptonic $B$-decays which are strictly related to the anomalies discussed so far. 
Since dominant effects occur at tree level, the inclusion of quantum effects is not relevant here.

\subsubsection{\texorpdfstring{$B \to \ell\nu$}{B to log}}
%%%%%%%%%%%%%%%%%%%%%%%%%%%%%%%%%%%%%%%%
%
A charged-current process closely related to $B \to D^{\mysmall (*)}\ell\nu$ is the decay $B \to \ell\nu$. We define the related LFUV observable, $R^{\tau/\ell}_{B\tau\nu}$, as
\begin{align}
R^{\tau/\mu}_{B\tau\nu} = 
\frac{\mathcal{B}(B \to \tau\nu)_{\rm exp}/\mathcal{B}(B \to \tau\nu)_{\rm SM}}{\mathcal{B}(B \to \mu\nu)_{\rm exp}/\mathcal{B}(B \to \mu\nu)_{\rm SM}} 
\approx 1 - \frac{2v^2}{\Lambda^2} \,  C_3 \,\lambda^{e}_{33}\left(\lambda^{d}_{33} + \frac{V_{us}\lambda^d_{23}\cos\gamma}{|V_{ub}|} \right)
\,,
\end{align}
where $\gamma \approx 70^\circ$.
Since Belle II aims to measure $R^{\tau/\mu}_{B\tau\nu}$ with a $5\%$ accuracy, it is likely that 
$R^{\tau/\mu}_{B\tau\nu}$ will provide a strong constraint to the present framework.

%%%%%%%%%%%%%%%%%%%%%%%%%%%%%%%%%%%%%%%%%%%%%%%%%%%%%%%%%%%
\subsubsection{\texorpdfstring{$B\to K^{\mysmall (*)} \nu\bar\nu$}{B to Knn}}
%%%%%%%%%%%%%%%%%%%%%%%%%%%%%%%%%%%%%%%%%%%%%%%%%%%%%%%%%%%

Another important process is $B \rightarrow K \bar \nu \nu$, which is strictly related to the neutral-current anomaly. 
%In fact, NP enters this process through the coefficients ${\cal C}^{\nu}_{\rm \mysmall NP}$ and ${\cal C}^{\nu \mysmall \prime}_{\rm \mysmall NP}$; as equation (\ref{eq:tree_matching}) shows, these are closely linked to the coefficients describing NP in $b \rightarrow s \ell^{\mysmall +} \ell^{\mysmall -}$, ${\cal C}^{\mysmall  9 (\prime)}_{\rm \mysmall NP}$ and  ${\cal C}^{\mysmall  10 (\prime)}_{\rm \mysmall NP}$. This happens because they ultimately originate from the same $SU(2)_{\mL} \times U(1)_{\mysmall Y}$ invariant operators in $\cL^{\mysmall 0}_{\rm \mysmall NP}$.\\
We consider the observable $R^{\nu\nu}_{K}$, defined as
\begin{equation}
R^{\nu\nu}_{K} = \frac{ \cB(B\to K \bar \nu \nu)}{\; \;\;\cB(B\to K\bar\nu \nu)_{\mysmall\rm SM}} \, ,
\end{equation}
which is subject to the experimental constraint $R_{K}^{\nu \nu} < 4.3$~\cite{Lutz:2013ftz,Buras:2014fpa}. In our framework $R^{\nu\nu}_{K}$ can be expressed as
\begin{equation}
R^{\nu\nu}_{K} = \frac{\sum_{ij} 
\left| {\cal C}^\nu_{ij} + {\cal C}^{\nu \mysmall \prime}_{ij} \right|^2}{3\left|{\cal C}_{\mysmall \rm SM}^\nu \right|^2}=\frac{\sum_{ij} 
\left| {\cal C}_{\mysmall \rm SM}^\nu \delta_{ij}+ ({\cal C}_{\rm \mysmall NP}^\nu)_{ij} + ({\cal C}_{\rm\mysmall NP}^{\nu^{\prime}})_{ij} \right|^2}{3\left|{\cal C}^{\nu}_{\mysmall \rm SM} \right|^2} \, .
\end{equation}
By expanding the numerator and using the property $\sum_{ij} \lvert \lambda^e_{ij} \rvert^2 =1$ and $\sum_i \lambda^e_{ii}=1$, we find
%{\setlength{\mathindent}{0cm} 
\begin{align}
R^{\nu\nu}_{K} \approx 1 + \frac23 \frac{\pi}{\alpha \left| {\cal C}_{\mysmall \rm SM}^\nu\lambda_{bs}^t \right|}\frac{v^2}{\Lambda^2} \left[ (C_1-C_3)\lambda^d_{23} +  C_4\Gamma^d_{23}\right]+ \frac{1}{3} \left( \frac{\pi}{\alpha \left| {\cal C}_{\mysmall \rm SM}^\nu\lambda_{bs}^t \right|}\frac{v^2}{\Lambda^2} \left[ (C_1-C_3)\lambda^d_{23} +  C_4\Gamma^d_{23}\right] \right)^2 \, .
\end{align}
%}
Since ${\cal C}_{\mysmall \rm SM}^\nu \approx -6.4$~\cite{Buras:2014fpa,Brod:2010hi}, we get the numerical result
%{\setlength{\mathindent}{0cm} 
\begin{align} 
R_{K}^{\nu \nu}\approx1 + 0.6\left( \frac{\lambda^d_{23}}{0.01} \frac{\left[ (C_1-C_3) +  C_4\Gamma^d_{23}/\lambda^d_{23}\right]}{\Lambda^2({\rm TeV}^2)}  \right)+ 0.3 \left( \frac{\lambda^d_{23}}{0.01} \frac{\left[ (C_1-C_3) +  C_4\Gamma^d_{23}/\lambda^d_{23}\right]}{\Lambda^2({\rm TeV}^2)} \right)^2 \, .
\end{align}
%}

\subsubsection{\texorpdfstring{$B_{s}\to\mu\bar\mu$}{Bs to mm}}

NP contributions for the observable $R^{\mu/e}_K$ can also enter the process $B_{s}\to\mu\bar\mu$. 
In particular, NP effects for $B_{s}\to\mu\bar\mu$ are encoded by the following expression
\begin{equation}
R_{B_s\mu\mu} = \frac{\mathcal{B}(B_s\to\mu\bar\mu)_{\rm exp}}{\mathcal{B}(B_s\to\mu\bar\mu)_{\rm \mysmall SM}} \simeq 
\left|\frac{{\cal C}_{\rm \mysmall SM}^{10} + ({\cal C}_{\rm \mysmall NP}^{10})_{\mu\mu} - ({\cal C}_{\rm \mysmall NP}^{10^{\prime}})_{\mu\mu}}{{\cal C}^{10}_{\rm \mysmall SM}}\right|^2, 
\label{eq:R}
\end{equation}
to be compared with the current experimental measurement and SM prediction 
for the branching ratio of this process~\cite{CMS:2014xfa,Bobeth:2013uxa}:
\begin{equation}
\mathcal{B}(B_s\to\mu\bar\mu)_{\rm exp}=2.8^{+0.7}_{-0.6}\times10^{-9}\qquad\qquad
\mathcal{B}(B_s\to\mu\bar\mu)_{\rm \mysmall SM}= 3.65(23)\times10^{-9}\,.
\end{equation}
%

%%%%%%%%%%%%%%%%%%%%%%%%%%%%%%%%%%%%%%%
\subsubsection{Lepton-flavour violating \texorpdfstring{\boldmath$B$\unboldmath}{B}\ decays}
\label{susec:H}
%%%%%%%%%%%%%%%%%%%%%%%%%%%%%%%%%%%%%%%

In our model, LFV decays like $B_s \to \tau^\pm \mu^\mp$ and $B\to K \tau^\pm \mu^\mp$ are generated at the tree level.
Their branching ratios are given by~\cite{Crivellin:2015era}
%
%\mathcal{B}\left(B_s \to \ell^\pm \ell^{\prime\mp} \right) &\simeq 4\times 10^{-8}\left| C_{9}^{\ell\ell^{\prime}}\right|^2
%
\begin{align}
\mathcal{B}\left(B_s \to \tau^\pm \mu^\mp \right) &\approx 2\times 10^{-8}
\left(
\left| 
({\cal C}_{\rm \mysmall NP}^{9})_{\tau \mu} - ({\cal C}_{\rm \mysmall NP}^{9^{\prime}})_{\tau \mu}
\right|^2  
+\left| 
({\cal C}_{\rm \mysmall NP}^{10})_{\tau \mu} - ({\cal C}_{\rm \mysmall NP}^{10^{\prime}})_{\tau \mu}
\right|^2
\right)
\nonumber\\
\mathcal{B}(B\to K \tau^\pm \mu^\mp) & \approx 2 \times 10^{-8} 
\left( 
\left| 
({\cal C}_{\rm \mysmall NP}^{9})_{\tau \mu} + ({\cal C}_{\rm \mysmall NP}^{9^{\prime }})_{\tau \mu}
\right|^2  
+\left| 
({\cal C}_{\rm \mysmall NP}^{10})_{\tau \mu} + ({\cal C}_{\rm \mysmall NP}^{10^{\prime }})_{\tau \mu}
\right|^2
\right)\,,
\label{bkstaumu}
\end{align}
where the factor of two in the above expressions accounts for the final state $\tau^\pm \mu^{\mp} = \tau^+ \mu^{-} + \tau^- \mu^{+}$.
As we will see shortly, loop-induced $\tau$ LFV decays are typically better probes of our scenario than LFV $B$-decays.

\subsection{One-loop phenomenology}

Electroweak corrections induce two main effects. First, $Z$ and $W$ couplings to fermions are modified with respect to the SM. 
Second, as we can see from eq.~(\ref{eq:effective_Lagrangian_GeV}) and related tables, a purely leptonic Lagrangian is also 
generated at low energies.  As a consequence, we expect LFV and LFUV effects in $Z$, $W$ and $\tau$ observables. 

\subsubsection{$Z$-pole observables} 

The NP modifications to $Z$ couplings arising in our setup, see eq.~(\ref{eq:Z_couplings_NP}), explicitly break both LFV and LFUV. 
The consequent deviations of $Z$-pole observables from SM expectations are tightly constrained by LEP measurements of the 
$Z$ decay widths, left-right and forward-backward asymmetries. We recall the definition of the axial and vector couplings
\begin{align}
v_{\ell}=(g_{\mL}^e)_{\ell \ell} + (g_{\mR}^e)_{\ell \ell} && a_{\ell}=(g_{\mL}^e)_{\ell \ell} - (g_{\mR}^e)_{\ell \ell} \, ,
\end{align}
and we consider the observables $v_\tau/v_e$ and $a_\tau/a_e$, which quantify the universality of $Z$ couplings to charged leptons. 
In our framework they read
\begin{align}
\begin{aligned}
\frac{v_\tau}{v_e} &\approx 1 - \frac{2}{1- 4 s_{\rm \mysmall W}^2} \left[ (\Delta g_{\mL}^e)_{33} -(\Delta g_{\mL}^e)_{11} +(\Delta g_{\mR}^e)_{33}-(\Delta g_{\mR}^e)_{11} \right]\\
\frac{a_\tau}{a_e} &\approx 1 - 2 \left[ (\Delta g_{\mL}^e)_{33} -(\Delta g_{\mL}^e)_{11} -(\Delta g_{\mR}^e)_{33}+(\Delta g_{\mR}^e)_{11} \right] \,,
\end{aligned}
\end{align}
leading to the following estimates
\begin{align} 
\begin{aligned}
\frac{v_\tau}{v_e} &\approx1-\frac{0.05}{\Lambda^2({\rm TeV}^2)}\left[(C_1 - C_3)\lambda^e_{33} + C_6\Gamma^e_{33} +0.2 C_3\lambda^e_{33} + 0.02 \left(( C_1 - C_4)\lambda^e_{33} + (C_6 -C_5)\Gamma^e_{33}\right) \right]  \\
\frac{a_\tau}{a_e} &\approx1-\frac{0.004}{\Lambda^2({\rm TeV}^2)}
\left[(C_1 - C_3)\lambda^e_{33} - C_6\Gamma^e_{33} +0.2 C_3\lambda^e_{33} + 0.02 \left(( C_1 - C_4)\lambda^e_{33} - (C_6 -C_5)\Gamma^e_{33}\right) \right]
\end{aligned}
\end{align}
to be compared with the LEP bounds~\cite{Agashe:2014kda}
\begin{align}
\frac{v_\tau}{v_e} = 0.959\; (29) && \frac{a_\tau}{a_e} = 1.0019\; (15) \, .
\label{eq:zpole_pdg}
\end{align}

Another important observable is the number of neutrinos $N_\nu$, which is extracted from the invisible $Z$ width. Taking the NP modification of  $Z$ couplings to neutrinos into account, $N_\nu$ can be approximated by 
\begin{align} 
N_\nu \approx 3+ 4 \sum_i(\Delta g_{\mL}^\nu)_{ii} \approx 3 + \frac{0.008}{\Lambda^2} \lambda^{e}_{33} \left[ \left(C_1 + C_3 \right) - 0.2 C_3 + 0.02 \left(C_1 - C_4\right)\right] \,,
\end{align}
while the experimental bound reads $N_\nu = 2.9840 \pm 0.0082$ \cite{Agashe:2014kda} . 
%\blue{Recentemente un collega mi ha fatto notare che estrarre il limite su $\Delta g_{\mL}^\nu$ dalla misura di $N_{\nu}$ non e' rigorosissimo, perche' $N_{\nu}$ sperimentalmente viene estratto assumendo che la Z accoppi universalmente ai leptoni carichi, cosa non vera nel nostro caso; per estrarre un limite su $\Delta g_{\mL}^\nu$ dai dati di LEP mi ha suggerito di far riferimento all'analisi 1503.07872v2. Non ho ancora riflettuto per bene su questa osservazione, ma intanto ve la faccio presente.}\\
Electroweak quantum corrections give rise also to the LFV decay mode $Z \to \mu^\pm \tau^\mp$. However, we have explicitly checked that in our model its branching ratio, typically of order $10^{-7}$, is always well below the current experimental bound $\mathcal{B}(Z \to \mu^\pm \tau^\mp)_{\rm exp} \leq 1.2 \times 10^{-5}$.
At the loop-level also the $W^\pm$ couplings to leptons are modified with respect to their SM expectations. 
However, the constraints on our model parameters arising from Z-pole observables are much stronger and therefore, 
hereafter, we neglect $W^\pm$ decays.

\subsubsection{Purely leptonic effective Lagrangian}
The effective low-energy Lagrangian (\ref{eq:effective_Lagrangian_GeV}) contains a purely leptonic Lagrangian $\cL_{\rm eff}^{\ell}$. 
Taking into account the explicit values of the $\xi_i$ and $\delta \xi_i$ for leptonic operators, and omitting terms manifestly respecting LFU in charged leptons,  we can write it as
\begin{align} 
\label{eq:purely_leptonic_Lagrangian}
{\cal L}_{\rm eff}^\ell= -\frac{4G_{\mysmall F}}{\sqrt{2}} & \Big[ (\bar e_{\mL} \gamma_\mu \lambda^e e_{\mL}) \sum_f (\bar f \gamma^{\mu} f) (2 g^f_{\rm \mysmall SM} c_t^e - Q_{\psi} c_{\gamma}^e) +  (\bar e_{\mR} \gamma_\mu \Gamma^e e_{\mR}) \sum_f (\bar f \gamma^\mu f) (2 g^f_{\rm \mysmall SM} c_t^{e \, \mysmall \prime}- Q_{f} c_{\gamma}^{e \, \mysmall \prime}) \nn  \\
&   + c_t^{cc} (\bar e_{\mL} \gamma_\mu \lambda^e \nu_{\mL}) (\bar \nu_{\mL} \gamma_\mu e_{\mL} + \bar u_{\mL} \gamma_\mu V_{\rm \mysmall CKM} d_{\mL}) + {\rm h.c.} \Big] \,  ,
\end{align}
where $f=\left\lbrace \nu_{\mL}, e_{\mL}, e_{\mR} \right\rbrace$ and $g^f_{\rm \mysmall SM}$ is the $Z$ coupling to the $f$ field in the SM. 
The coefficients $c_t^e$, $c_{\gamma}^e$, $c_t^{e \, \mysmall \prime}$, $c_{\gamma}^{e \, \mysmall \prime}$ are given by
\begin{align}
&c_t^e=\frac{3 v^2 y_t^2}{32 \pi^2 \Lambda^2}  (C_1 - C_3) \lambda^u_{33} \log \frac{\Lambda^2}{m_{\rm \mysmall EW}^2} \nn \\
&c_t^{e^{\prime}}=\frac{3 v^2 y_t^2}{32 \pi^2 \Lambda^2}   C_6 \lambda^u_{33} \log \frac{\Lambda^2}{m_{\rm \mysmall EW}^2} \nn \\
&c_{\gamma}^e=\frac{v^2 e^2}{48 \pi^2 \Lambda^2}  \left[ (3C_3 - C_1 + C_4) \log \frac{\Lambda^2}{\mu^2} + 2 (C_1 - C_3) \left( \lambda^u_{33} \log \frac{m_{\rm \mysmall EW}^2}{\mu^2} + \lambda^u_{22} \log \frac{m_c^2}{\mu^2}\right) \right. \nn \\
& \left. \qquad \qquad  \qquad- \left((C_1 + C_3)\lambda^d_{33}  + C_4\Gamma^d_{33}  \right) \log \frac{m_b^2}{\mu^2} \right] \nn \\
&c_{\gamma}^{e^{\prime}}=\frac{v^2 e^2}{48 \pi^2 \Lambda^2}  \left[ (C_6 - C_5) \log \frac{\Lambda^2}{\mu^2} + 2 C_6 \left( \lambda^u_{33} \log \frac{m_{\rm \mysmall EW}^2}{\mu^2} + \lambda^u_{22} \log \frac{m_c^2}{\mu^2}\right) - \left(C_6\lambda^d_{33}  + C_5\Gamma^d_{33} \right)\log \frac{m_b^2}{\mu^2}\right] \nn \\
&c_t^{cc}=\frac{3 v^2 y_t^2}{16 \pi^2 \Lambda^2}  C_3 \lambda^u_{33} \log \frac{\Lambda^2}{m_t^2} \,\,  .
\end{align}
%\end{empheq}
%
Notice that, in all observables analysed in this work but $R^{\tau/\ell_{1,2}}_\tau$ (see eq.~\ref{eq:def_R_tau}), 
we systematically neglected corrections to the Fermi constant. Their inclusion would amount to replace 
$G^0_F = v^2/\sqrt{2}$ with $G^0_F \simeq G_F(1-c^{cc}_t \lambda^{e}_{22})$ where $G_F$ is the value extracted from the muon decay rate measurement. Numerically,
such correction is below the $0.1\%$ level and therefore safely negligible
since $G^0_F \approx G_F( 1 - 0.004 \, \lambda^e_{22}\,C_3/\Lambda^2({\rm TeV}))$ with $\lambda^e_{22} \ll 1$.

Lagrangian (\ref{eq:purely_leptonic_Lagrangian}) manifestly generates both LFV and LFUV processes. Given the hierarchy in $\lambda^e_{ij}$ and $\Gamma^e_{ij}$, NP effects are maximized in transitions involving the third generation. As a consequence, we focus on $\tau$ decays such as
$\tau \rightarrow \ell \bar \nu \nu$ and $\tau \rightarrow 3 \mu$.

\subsubsection{$\tau \to \ell \bar\nu \nu$}
\noindent LFU breaking effects in $\tau \to \ell \bar\nu \nu$ (with $\ell_{1,2}=e,\mu$) are described by the observables
\begin{align}
R^{\tau/\ell_{1,2}}_\tau = \frac{\mathcal{B}(\tau \to \ell_{2,1} \nu\bar\nu)_{\rm exp}/\mathcal{B}(\tau \to \ell_{2,1} \nu\bar\nu)_{\rm \mysmall SM}}{\mathcal{B}(\mu \to e \nu\bar\nu)_{\rm exp}/\mathcal{B}(\mu \to e \nu\bar\nu)_{\rm \mysmall SM}} \,,
\label{eq:def_R_tau}
\end{align}
which are subject to the strong experimental constraints $R^{\tau/\mu}_\tau = 1.0022 \pm 0.0030$ and $R^{\tau/e}_\tau = 1.0060 \pm 0.0030$~\cite{Pich:2013lsa}. Taking into account the correlation of these measurements, the combined constraint reads
\begin{align}
R^{\tau/\ell}_\tau = 1.0032 \pm 0.0026\,.
\label{eq:tau_LFU_data}
\end{align}
In our setup the effective Lagrangian describing $e_{\alpha} \rightarrow e_{\beta} \bar \nu_j \nu_i$ is given by
\begin{align}
{\cal L}= - \frac{4 G_{\mysmall F}}{\sqrt{2}} 
\left[
({\cal C}^{\alpha \beta}_{\mL})_{ij}(\bar e_{\beta \mL} \gamma_\mu e_{\alpha \mL} )(\bar \nu_{i \mL} \gamma^\mu \nu_{j \mL}) + ({\cal C}^{\alpha \beta}_{\mR})_{ij} (\bar e_{\beta \mR} \gamma_\mu e_{\alpha \mR} )(\bar \nu_{i \mL} \gamma^\mu \nu_{j \mL})  
\right] \,  ,
\end{align}
where 

\begin{align}
({\cal C}^{\alpha \beta}_{\mL})_{ij} &=\delta_{\beta j} \delta_{i \alpha} + c_t^e \delta_{ij} \lambda^e_{\beta \alpha} + c_t^{cc}(\lambda^e_{\beta j} \delta_{i \alpha}+\lambda^e_{i \alpha} \delta_{\beta j})\,,
\label{eq:C_L} 
\\
({\cal C}^{\alpha \beta}_{\mR})_{ij} &=c_t^{e^{\prime}} \Gamma^e_{\beta \alpha} \delta_{ij}\,.
\label{eq:C_R} 
\end{align}
Notice that the SM contribution to $e_{\alpha} \rightarrow e_{\beta} \bar \nu_j \nu_i$ is accounted for by the first term of (\ref{eq:C_L}).\\
The ratio $R^{\tau/\ell}_\tau$ can be expressed in terms of these coefficients as follows
\begin{equation}
R^{\tau/\ell_{1,2}}_{\tau} = \frac{\sum_{ij} |({\cal C}^{\tau \; \ell_{\mysmall 2,1}}_{\mL})_{ij}|^2 + | ({\cal C}^{\tau\; \ell_{\mysmall 2,1}}_{\mR})_{ij} |^2}{\sum_{ij} |({\cal C}^{\mu e}_{\mL})_{ij} |^2 +|({\cal C}^{ \mu e}_{\mR})_{ij} |^2} \,  .
\label{eq:R_D}
\end{equation}
Working linearly in the NP contribution, we find that
\begin{align}
R^{\tau/e}_\tau &\simeq 1 + 2\,c^{cc}_t \lambda^{e}_{33}
\approx 1 + 0.008 \,  \lambda^e_{33}\frac{C_3}{\Lambda^2({\rm TeV})}\nn\\
R^{\tau/\mu}_\tau &\simeq 1 + 2\,c^{cc}_t  (\lambda^e_{33}-\lambda^e_{22})
\approx 1 + 0.008 \,  (\lambda^e_{33}-\lambda^e_{22})\frac{C_3}{\Lambda^2({\rm TeV})} \,  .
\label{eq:tau_LFU}
\end{align}

\subsubsection{$\tau \rightarrow 3 \mu$}
\noindent One of the most studied LFV processes generated by $\cL_{\rm eff}^\ell$ is the decay $\tau \rightarrow 3\mu$, which is forbidden in the SM. 
The only contribution is given by $\cL_{\rm eff}^\ell$ 
\begin{align}
\cL_{\rm eff}^\ell=-\frac{4 G_{\mysmall F}}{\sqrt{2}} &
\left\{\lambda^e_{23} \left[  (c_{\mysmall LR} - c_t^e)(\mu_{\mL} \gamma_\mu \tau_{\mL})(\bar \mu_{\mL} \gamma^\mu \mu_{\mL}) + c_{\mysmall LR}(\mu_{\mL} \gamma_\mu \tau_{\mL})(\bar \mu_{\mR} \gamma^\mu \mu_{\mR}) \right] \right.\nn 
\\
& \left.+\Gamma^e_{23} \left[  (c_{\mysmall LR}^{ ^{\prime}} - c_t^{e^{\prime}})(\mu_{\mR} \gamma_\mu \tau_{\mR})(\bar \mu_{\mL} \gamma^\mu \mu_{\mL}) + c_{\mysmall LR}^{^{\prime}}(\mu_{\mR} \gamma_\mu \tau_{\mR})(\bar \mu_{\mR} \gamma^\mu \mu_{\mR}) \right]\right\} + \dots \, ,
\end{align}
where $c_{\mysmall LR}^{ ^{(\prime)}}=2 s_{\rm \mysmall W}^2 c_t^{e^{(\prime)}} + c_\gamma^{e^{(\prime)}} $. Adapting the formula given in ref. \cite{Kuno:1999jp} we find
\begin{align}
\Gamma(\tau \rightarrow 3 \mu)= \frac{G_{\mysmall F}^2 m_\tau^5}{192 \pi^3} \left\{\left[ 2 (c_{\mysmall LR} -c_t^e)^2 + c_{\mysmall LR}^2 \right]\left| \lambda^e_{23}\right|^2 +\left[ 2c_{\mysmall LR}^{^{\prime} \, 2}  + (c_{\mysmall LR}^{^{\prime}}-c_t^{e^{\prime}})^2  \right]\left| \Gamma^e_{23}\right|^2\right\} \, .
\end{align}
Keeping only the Yukawa contribution, which is typically the dominant one, we end up with the following numerical estimate
\begin{align}
{\cal B}(\tau \rightarrow 3 \mu) \approx
\left(\frac{\lambda^e_{23}}{0.3}\right)^2\left[5.0  \frac{(C_1-C_3)^2}{\Lambda^4 ({\rm TeV}^4)} +4.5  \frac{C_6^2}{\Lambda^4({\rm TeV}^4)}\left(\frac{\Gamma^e_{23}}{\lambda^e_{23}}\right)^2\right] \cdot 10^{-8}\,,
\end{align}
to be compared with the current experimental bound ${\cal B}(\tau \rightarrow 3 \mu) \leqslant 1.2 \cdot 10^{-8}$ \cite{Amhis:2016xyh}.

%%%%%%%%%%%%%%%%%%%%%%%%%%%%%%%%%%%%%%%%%%%%%%%%%%%%%%%%%%%%%%%%%%
\subsection{Numerical analysis}
%%%%%%%%%%%%%%%%%%%%%%%%%%%%%%%%%%%%%%%%%%%%%%%%%%%%%%%%%%%%%%%%%%
\label{sec:scenario}
\noindent In this section, we focus on a phenomenologically relevant scenario where only $(\cC^{\mysmall 9}_{\rm \mysmall NP})_{\mu \mu}$ 
is non-vanishing. This can be achieved by imposing the following conditions
\begin{align}
&\Gamma^e_{ij}=\lambda^e_{ij}&&C_1 + C_3 = C_6  && C_4=C_5=0 \, .
\label{eq:conditions}
\end{align}
Taking the NP scale to be $\Lambda \approx 1\; {\rm TeV}$, the free parameters in this setup are $C_1$, $C_3$,  $\lambda^d_{23}$ 
and $\lambda^e_{23}$ where $\lvert \lambda^{e,d}_{23} \rvert \leq 0.5$~\cite{Feruglio:2016gvd,Feruglio:2017rjo}.
We can further restrict the bounds on $\lambda^e_{23}$ because the non-observation of LFUV in $R^{\mu/e}_D$ implies that
$\left| \lambda^e_{22} \right| \approx \left| \lambda^e_{23} \right|^2 \leq 0.1$ \cite{Greljo:2015mma}.
As to $C_{1,3}$, we assume $\left| C_{1,3}\right| \leq 3$.
Given (\ref{eq:conditions}), we obtain the following expressions for $B$-physics observables
\begin{align}
&R_{D^{\mysmall(*)}}^{\tau / \ell}=1-0.12\;\frac{C_3}{\Lambda^2({\rm TeV}^2)} \lambda^e_{33} \left(\frac{V_{cs}}{V_{cb}} \lambda^d_{23} + \lambda^d_{33}\right) \nn \\
&R_{K}^{\mu / e}=1-\frac{0.30}{\Lambda^2({\rm TeV}^2)} \frac{\lambda^e_{22}\lambda^d_{23}}{10^{-3}} \left( C_1 + C_3\right) \nn \\ 
&R_{K^{\mysmall *}}^{\mu / e}=1-\frac{0.23}{\Lambda^2({\rm TeV}^2)}\frac{\lambda^e_{22} \lambda^d_{23}}{10^{-3}} 
\left( C_1 + C_3\right) \nn \\ 
&R_K^{\nu \nu}=1 + 0.6\left( \frac{\lambda^d_{23}}{0.01} \frac{C_1 -C_3}{\Lambda^2({\rm TeV}^2)}   \right)+ 0.3 \left( \frac{\lambda^d_{23}}{0.01} \frac{C_1 -C_3}{\Lambda^2({\rm TeV}^2)}   \right) ^2 \, .
\end{align}
On the other hand, Z-pole observables simplify to
\begin{align} 
&\frac{v_\tau}{v_e} =1-\frac{0.05\, \lambda^e_{33}}{\Lambda^2({\rm TeV}^2)}(2\;C_1+0.2 \; C_3 + 0.02 \; (2 \; C_1 + C_3) )\nn \\
&\frac{a_\tau}{a_e} =1+  0.007 \,  \lambda^e_{33} \frac{C_3}{\Lambda^2({\rm TeV}^2)}  \nn \\
&N_\nu=3 + \frac{0.008 \lambda^e_{33}}{\Lambda^2({\rm TeV}^2)} ( C_1 + C_3  - 0.2\;  C_3 + 0.02 \; C_1) \,.
\end{align}
\begin{figure}[H]
\centering
\includegraphics[scale=0.12]{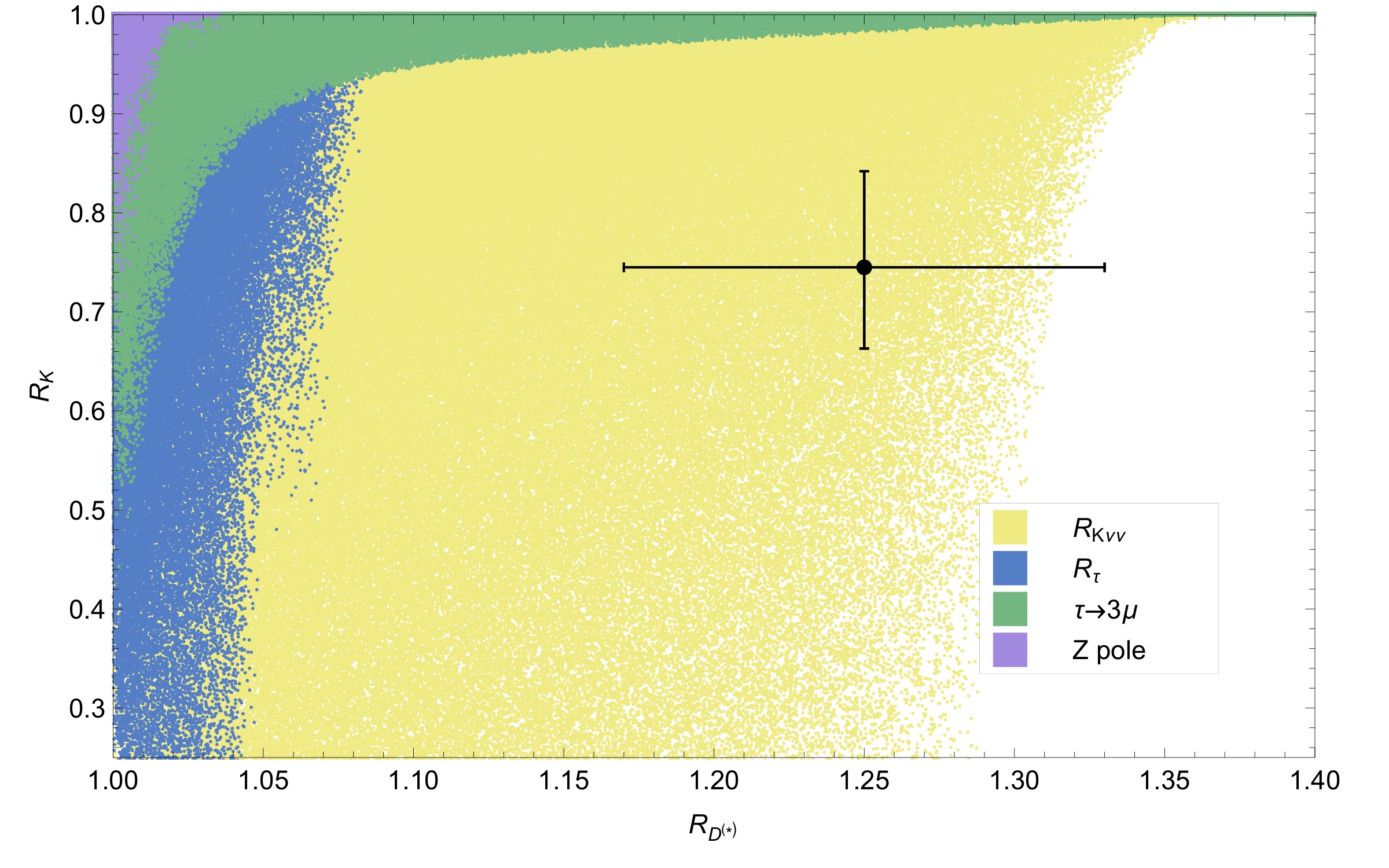}
\caption{Impact of one-loop-induced constraints on the values of $R_{D^{\mysmall (*)}}^{\tau/\ell}$ and $R_{K}^{\mu/e}$ for  $C_1 \in \{-3,3\}$, $C_3 \in \{-3,3\}$,  $\lambda^e_{23} \in \{-0.3,0.3\}$ and $\lambda^d_{23} \in \{-0.04,0.04\}$ and $\Lambda=1\,$TeV. }
\label{fig:anomalies}
\end{figure}
Finally, for $\tau$ decays, we obtain the following estimates
\begin{align}
&R_\tau^{\tau / \ell_{1,2}} = 1 + 0.008\, \lambda^e_{33} \frac{C_3}{\Lambda^2({\rm TeV}^2)} \nn\\
&\cB(\tau \rightarrow 3 \mu)= \left(\frac{\lambda^e_{23}}{0.3}\right)^2
\left[ 5.0  \frac{(C_1-C_3)^2}{\Lambda^4({\rm TeV}^4)} +4.5 \frac{(C_1 + C_3)^2}{\Lambda^4({\rm TeV}^4)} \right] \cdot 10^{-8} \, .
\end{align}
It is interesting to observe that the ratio $a_\tau/a_e$ depends exclusively on the Wilson coefficient $C_3$ of the charged-current operator. 
Choosing $ \lvert \lambda^d_{23} \rvert \lesssim V_{cb} $ in order to avoid too much fine tuning when reproducing the CKM matrix, there is 
a strong correlation among $R_{D^{\mysmall (*)}}^{\tau/\ell}$, $a_\tau/a_e$ and $R_\tau^{\tau / \ell_{1,2}}$. 
In particular, it turns out that the NP room left to $R_{D^{\mysmall (*)}}^{\tau/\ell}$ is significantly reduced after taking into account all existing bounds.
This can be clearly seen in the graph displayed in fig.~\ref{fig:anomalies}, which shows the allowed regions for 
$R_{K}^{\mu/e}$ and $R_{D^{(*)}}^{\tau/\ell}$ after imposing the experimental bounds on Z-pole and $\tau$ observables at $2 \sigma$ level\footnote{We do not show the plot in the $R_{K^{\mysmall *}}^{\mu/e}$ vs. $R_{D^{(*)}}^{\tau/\ell}$ plane since it is almost indistinguishable
to that of fig.~\ref{fig:anomalies}.}.
Altough all observables receiving NP contribution at one loop impose strong bounds on $B$ anomalies, $Z$-pole observables set the stringest limits, forcing $\delta R_{D^{\mysmall (*)}}^{\tau/\ell}$ to be $ \lesssim 0.05$. Like in \cite{Feruglio:2016gvd, Feruglio:2017rjo}, we conclude that current data on $\tau$ and $Z$-pole observables challenge a simultaneous explanation of the present values of $R_{K^{\mysmall(*)}}^{\mu/e}$ and $R_{D^{\mysmall (*)}}^{\tau/\ell}$, when NP above the electroweak scale mainly affects the operator ${\mathcal O}_9$ and the third generation. 

\begin{figure}[H]
\centering 
\includegraphics[scale=0.6]{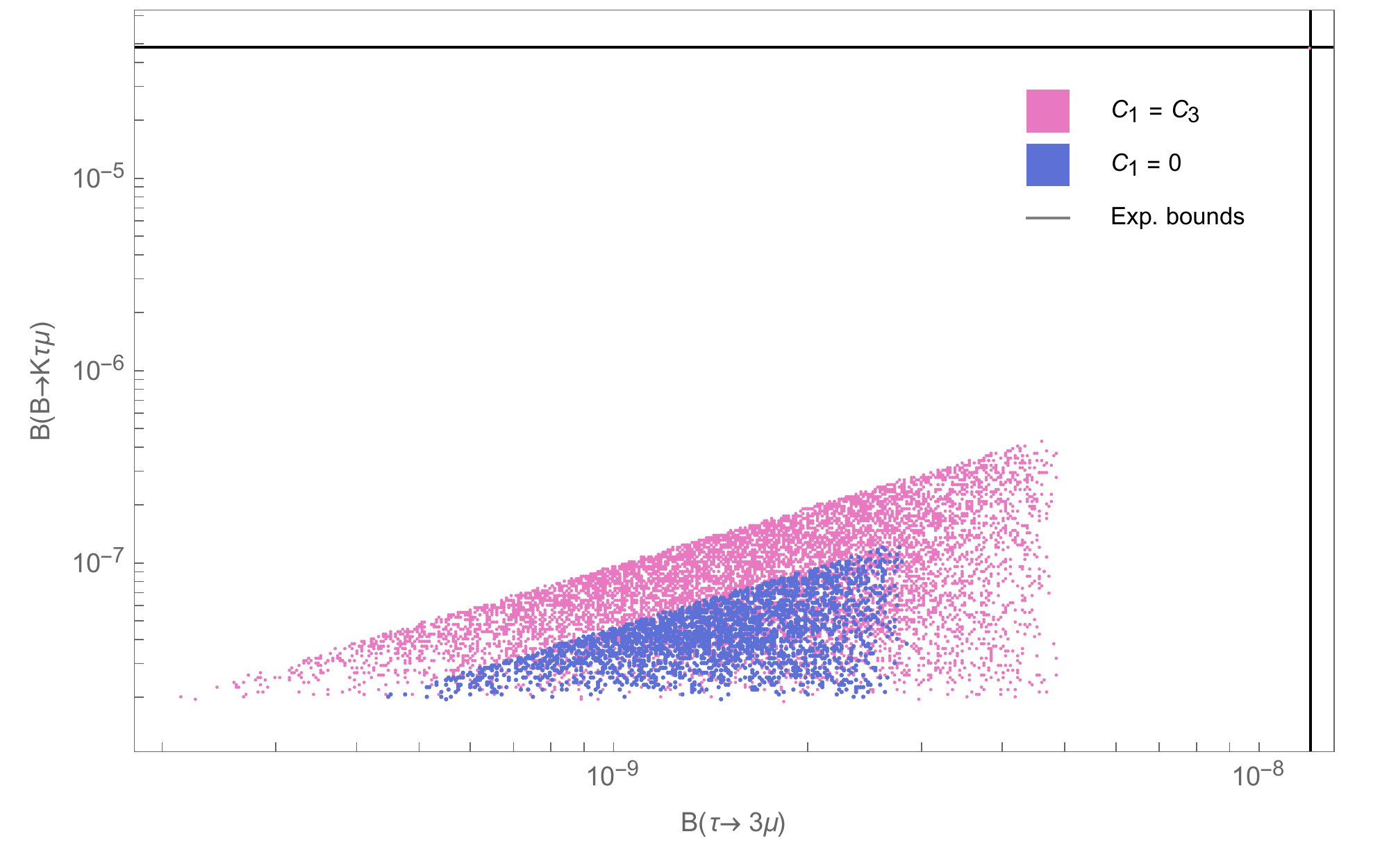}
\caption{$\cB(B \rightarrow 3 \mu)$ vs. $\cB(B \rightarrow K \tau \mu)$ within our model for two different configurations of $C_1$, $C_3$, imposing 
all constraints but $R_{D^{(*)}}^{\tau/\ell}$. We let parameters vary in the ranges 
$C_1 \in \{-3,3\}$, $C_3 \in \{-3,3\}$,  $\lambda^e_{23} \in \{-0.3,0.3\}$ and $\lambda^d_{23} \in \{-0.04,0.04\}$ and $\Lambda=1\,$TeV.}
\label{fig:ktau}
\end{figure}

In the plot of fig.~\ref{fig:ktau} we analyse the correlation between the branching ratios of LFV decays, $B \rightarrow K \tau \mu$ and 
$\tau \rightarrow 3 \mu$. The graph shows that the loop-induced process $\tau \rightarrow 3 \mu$ is a much more sensitive probe of the 
considered scenario than the tree level observable $B \rightarrow K \tau \mu$, due to the current and expected future experimental resolution.   

\section{Conclusions}

The persisting and coherent anomalous data in semileptonic B-decays point towards New Physics scenarios with 
large sources of Lepton Flavour Universality Violation. If this is the case, one would expect other non-standard 
effects to show up in low- and/or high-energy observables. The experimental signatures of specific scenarios 
able to accommodate these anomalies have been discussed extensively in the recent literature.
On the other hand, the importance of including electroweak corrections in scenarios with left-handed semileptonic 
operators defined at the scale $\Lambda \gg v$ was stressed in Ref.~\cite{Feruglio:2016gvd,Feruglio:2017rjo}.

In this work, by assuming that New Physics mainly affects the third generation, we have generalised the analysis of~\cite{Feruglio:2016gvd,Feruglio:2017rjo} 
by considering an effective theory involving both purely left-handed operators 
$(V-A)\times(V-A)$ and operators with right-handed currents of the form $(V + A)\times(V + A)$ and $(V \pm A)\times(V \mp A)$.
In this framework, we have derived the low-energy effective Lagrangian by means of the running and matching procedure outlined in~\cite{Feruglio:2016gvd,Feruglio:2017rjo}.
As in the previous analysis,  we find that the dominant effects concern the corrections to the leptonic couplings of the $W$ and $Z$ 
vector bosons as well as the generation of a purely leptonic effective Lagrangian. 
Then we focused on a phenomenologically favoured setup where the dominant New Physics effects are encoded in the low-energy 
Wilson coefficient $C_9$~\cite{Altmannshofer:2014rta,Descotes-Genon:2015uva,Hiller:2014yaa}.
As our numerical analysis shows, also in this case the inclusion of electroweak corrections are mandatory to obtain reliable 
predictions.
In particular, we confirm and reinforce the conclusion that the stringent experimental bounds on Z-pole observables and $\tau$ 
decays severely reduce the New Physics room for a simultaneous explanation of charged and neutral-current non-standard data.

%%%%%%%%%%%%%%%%%%%%%%%%%%%%%%%%%%%%%%%
\section*{Acknowledgements}

We thank David Straub and Olcyr Sumensari for useful discussions. 
This work was supported in part by the MIUR-PRIN project 2010YJ2NYW and by the European Union network FP10 ITN ELUSIVES and INVISIBLES-PLUS (H2020- MSCA- ITN- 2015-674896 and H2020- MSCA- RISE- 2015- 690575).
The research of C.C. was supported in part by the Swiss National Science Foundation (SNF) under contract 200021-159720.
The research of P.P. was supported in part by the ERC Advanced Grant No.  267985 (DaMeSyFla), by the research grant TAsP, and by the INFN.

\end{document}